\documentclass[journal=jcisd8,manuscript=article,email=true,dvipsnames,usenames,maxauthors=0]{achemso}

\usepackage{chemformula} 
\usepackage[T1]{fontenc} 

\usepackage{soul}
\usepackage[normalem]{ulem}

\SectionNumbersOn

\newcommand{\marked}[1]{}

\newcommand{\mathnewydd}[1]{#1}
\newcommand{\newydd}[1]{#1}

\newcommand{\kbd}[1]{\texttt{\textcolor{RawSienna}{#1}}}
\newcommand{\lnk}[1]{\texttt{\textcolor{NavyBlue}{\url{#1}}}}
\newcommand{\ttl}[1]{\texttt{\textcolor{ForestGreen}{#1}}}  
\newcommand{\tttl}[1]{\texttt{\textcolor{ForestGreen}{#1}}}  

\newcommand{\cf}{\latin{cf.}}
\newcommand{\eg}{\latin{e.g.}}
\newcommand{\etal}{\latin{et al.}}
\newcommand{\etc}{\latin{etc.}}
\newcommand{\ie}{\latin{i.e.}}
\newcommand{\nb}{\latin{n.b.}}

\author{Martin Thomas Horsch}
\email{martin.horsch@stfc.ac.uk}
\affiliation[STFC]{\small STFC Daresbury Laboratory, UKRI, Keckwick Ln, Daresbury, Cheshire WA4 4AD, UK}
\author{Christoph Niethammer}
\email{niethammer@hlrs.de}
\affiliation[HLRS]{\small High Performance Computing Center Stuttgart, Nobelstr.~19, 70569 Stuttgart, Germany}
\author{Gianluca Boccardo}
\affiliation[POLITO]{\small Department of Applied Science and Technology, Institute of Chemical Engineering, Politecnico di Torino, Corso Duca degli Abruzzi 24, 10129 Torino, Italy}
\author{Paola Carbone}
\affiliation[UoM]{\small School of Chemical Engineering and Analytical Science, University of Manchester, Oxford Rd, Manchester M13 9PL, UK}
\author{Silvia Chiacchiera}
\affiliation[STFC]{\small STFC Daresbury Laboratory, UKRI, Keckwick Ln, Daresbury, Cheshire WA4 4AD, UK}
\author{Mara Chiricotto}
\affiliation[UoM]{\small School of Chemical Engineering and Analytical Science, University of Manchester, Oxford Rd, Manchester M13 9PL, UK}
\author{Joshua D.\ Elliott}
\affiliation[UoM]{\small School of Chemical Engineering and Analytical Science, University of Manchester, Oxford Rd, Manchester M13 9PL, UK}
\author{Vladimir Lobaskin}
\affiliation[UCD]{\small School of Physics, University College Dublin, Dublin 4, Ireland}
\author{Philipp Neumann}
\affiliation[HSUHH]{\small \marked{\sout{Department of Informatics, Universit\"at Hamburg, Bundesstr.~45a, 20146}~}\newydd{High Performance Computing, Helmut-Schmidt-Universit\"at, Holstenhofweg 85, 22043} Hamburg, Germany}
\author{Peter Schiffels}
\affiliation[IFAM]{\small Fraunhofer Institute for Manufacturing Technology and Advanced Materials, Wiener Str.~12, 28359 Bremen, Germany}
\author{Michael A.\ Seaton}
\author{Ilian T.\ Todorov}
\affiliation[STFC]{\small STFC Daresbury Laboratory, UKRI, Keckwick Ln, Daresbury, Cheshire WA4 4AD, UK}
\author{Jadran Vrabec}
\affiliation[TU Berlin]{\small Thermodynamics and Process Engineering, Technische Universit\"at Berlin, Ernst-Reuter-Platz 1, 10587 Berlin, Germany}
\author{Welchy Leite Cavalcanti}
\affiliation[IFAM]{\small Fraunhofer Institute for Manufacturing Technology and Advanced Materials, Wiener Str.~12, 28359 Bremen, Germany}

\title[Semantic interoperability]{Semantic interoperability and characterization of data provenance in computational molecular engineering}

\keywords{Simulation workflow, data provenance, semantic interoperability, ontology, data management}

\begin{document}
\newpage
\begin{abstract}
By introducing a common representational system for metadata
that describe the employed simulation workflows,
diverse sources of data and platforms in computational
molecular engineering, such as workflow management systems, can
become interoperable at the semantic level. To achieve semantic
interoperability, the present work introduces two ontologies that provide a formal specification
of the entities occurring in a simulation workflow and the relations between them:
The software ontology VISO is developed to represent software packages and
their features, and OSMO, an ontology for simulation, modelling, and
optimization, is introduced on the basis of MODA, a previously developed
semi-intuitive graph notation for workflows in materials modelling.
As a proof of concept, OSMO is employed to describe a\marked{~\sout{leading-edge
application scenario}} \newydd{use case} of the TaLPas workflow management system\newydd{,
a scheduler and workflow optimizer for particle-based simulations}.
\end{abstract}

\section{Introduction}
\label{sec:intro}


Hans Hasse, to whose achievements this special issue is dedicated,
is among those who have contributed to the success of modelling and
simulation by computational molecular engineering. Building on previous
efforts in molecular model characterization and simulation method development,
\eg, by M\"oller and Fischer\citep{MF90} as well as Lotfi \etal,\citep{LVF92}
Vrabec and Hasse introduced the grand equilibrium
simulation method by which vapour-liquid equilibria can be efficiently
sampled.\citep{VH02} This workflow, implemented
in the \textit{ms2} code,\citep{DESLGGMBHV11, GRRDKGWHBWHV14, RKGJSGBWSKRDHHV17}
was the basis for a period of increased productivity in molecular model design
during which Hasse, Vrabec, and collaborators parameterized a
multitude of reliable intermolecular pair
potentials\citep{VSH01, EVH08a, SVH08, EMVH11, HHHV11, SHVH19} and applied them
to predict the thermodynamic properties
of pure components and mixtures.\citep{EVH08a, GNVH08, HVH09, PGHV13, WKLHHH16}
Using their code \textit{ls1 mardyn},\citep{NBBBEHWBGHVH14}
a molecular dynamics (MD) system size world record with four
trillion particles was achieved,\citep{EHBBHHKVHHBGNBB13}
which has recently been pushed towards
twenty trillion particles.\citep{TSHVGHBGNHKRKHBN19}
This work in model and software development, in combination with
the increase in accessible computational resources,
\marked{\sout{has} }played a role in establishing molecular modelling
as a branch of simulation-based engineering.

Improving the interoperability
of methods and codes, providing simulation metadata in an agreed way,
and specifying simulation workflows that integrate multiple model granularity
levels have become key challenges at combining computational
molecular engineering with the other simulation-based
engineering approaches that are already
widespread in industrial practice, \eg, computational
fluid dynamics (CFD) and process simulation. This requires a coordinated
effort in data technology. With this perspective, Burger,
von Harbou, and Hasse, jointly with industrial partners,
worked towards interfacing experimental and simulation data with
model design,\citep{AB4WBHKH14, AB6HKH15}
an objective that the ongoing virtual marketplace initiatives
promise to pursue systematically.\citep{Cavalcanti19, Hashibon19}
Within \newydd{a collaborative research centre (}CRC 926 \newydd{MICOS)},
Hasse and collaborators introduced the\marked{~\sout{OMEB}}
concept\marked{~\sout{for}} \newydd{of} domain-specific
processing-morphology-property relationships \newydd{for
component surfaces, referred to as OMEB from the German
expression.\citep{ASMKH16} This facilitates}\marked{~\sout{facilitating}} an
approach that can be employed to connect
molecular and phenomenological modelling to decision support by
multicriteria optimization,\citep{BH13, BBABBNSWKH14, BBHKSABKH17, FBLHB18} translating
problems of industrial end users to solutions based on
quantitatively reliable modelling and simulation.\citep{WKLHHH16, HRSBH17, FJBBVH19}
Recent works by Hasse and Lenhard address the philosophy of
modelling, formulating an engineering-oriented perspective
on the role of computational methods.\citep{HL17, LH17}
These contributions have advanced data technology in
materials modelling and created opportunities
to address further challenges, some of which\marked{~\sout{will be}}
\newydd{are} discussed in the present work.\marked{\newydd{~(Note: Following
a suggestion by one of the Reviewers, the preceding
two paragraphs were moved forward.)}}

Where databases and platforms using different data structures and file
formats interoperate, or where data and metadata from various sources are combined,
agreement on semantics becomes a necessity,\citep{SI19}
\newydd{supporting the effort to construct a universal
web into which any linked data can be integrated\citep{MJ11} to
become FAIR, \ie, findable, accessible,
interoperable, and reusable.\citep{Bicarregui16, Mons18} For this
purpose, metadata (\ie, data about data) need to be provided to
characterize the context of any relevant data items so that they
can be found and accessed easily and reused properly. This includes
information on data provenance, \ie, the process by which
the data have been obtained.\citep{MMCMGP17, SI19} Interoperability
applies to three major aspects of data stewardship:\citep{Mons18} Syntax (formats),
semantics (meaning), and pragmatics (procedures). It is achieved by
establishing a common intermediate standard to which all users and contributors
to a data infrastructure can map their own internal
approach.\citep{ZJDW18, LSEST19}}\marked{~\sout{Therefore}~}\newydd{Accordingly},
data technology solutions that aim at facilitating interoperability and
data integration require the definition of semantic
assets, \ie, documents that codify semantics.\citep{MWT11}
For this purpose, it is crucial to develop and
maintain community-governed semantic standards, facilitating
the systematic annotation of pre-existing dark data, \ie, data for which
machine-processable metadata are absent or insufficient,\citep{Heidorn08} by a
variety of data and metadata owners and infrastructure
providers. \newydd{In data technology, an ontology is a formal machine-processable
representation of knowledge within a certain domain. It consists
of a definition of classes, individuals (objects that belong to the classes),
and rules for the possible relations between them:\citep{MWM08, AH11, MAD17, LAL19} For instance, concerning
simulation workflows in materials modelling, a ``workflow node''
can be defined as a ``workflow graph that contains
exactly one workflow resource.'' Therein, \textit{workflow node}, \textit{workflow
graph}, and \textit{workflow resource} are classes, and \textit{contains} is a relation.
In this way, an ontology can be employed to standardize the semantic
space belonging to a particular application domain.}

\marked{\sout{In particular,}~}A variety of applications in simulation based engineering
can benefit from a machine-readable way of
specifying a simulation workflow;\citep{PCSMK16} thereby, the
characterization of workflows is relevant in two major ways. First,
workflows are designed and communicated within simulation environments
where materials models are evaluated to generate data by simulation.\citep{RC07}
Second, in order to integrate data obtained in different
ways (\eg, from simulation and experiment, or from simulations
with different models or solvers), simulation results need to be
stored together with metadata that describe their provenance\marked{\sout{,
\ie, the process by which they have been produced}}.
If experimental and other data are meant to be
integrated with simulation results in a common infrastructure,\citep{AB4WBHKH14, AB6HKH15}
workflow descriptions can be combined with domain-specific
provenance description ontologies which, \eg, already exist
in genetics\citep{BSH16} and nanosafety.\citep{HJOTMSW15}
Specified workflow metadata, supplemented by an extensive technical documentation,
can be employed to reproduce data by repeating the same workflow. Furthermore, certain aspects
of the data uncertainty (and uncertainty propagation),
such as the sensitivity with respect to specific model parameters
or the choice of a particular solver implementation, can be quantified by varying
individual values, parameters, or elements of a workflow;\cite{BABB17} \eg, round-robin studies
can be conducted, where various simulation software environments
are employed to carry out the same (or closely related)
algorithms in combination with the same
models, comparing the outcome.\citep{SMKBKSGRRKLGLVH17}
Other technologies that can profit from well-defined workflow semantics
include high performance computing (HPC) and scheduling environments where computational requirements
may be automatically predicted\cite{SVW19} and optimized by workflow autotuning and
task-based parallelization.\cite{CE19}

In computational molecular engineering,
two major organized efforts toward achieving an agreed coherent
semantic-technology framework have been conducted: With a focus
on process simulations, CAPE-OPEN was developed \newydd{in \textit{computer
aided process engineering} as an \textit{open} interface
standard.\citep{BP02, BP14} CAPE-OPEN}\marked{~\sout{which}} is in use both in
academic and industrial engineering practice.\citep{LCZ09, KTC17}
At present, ongoing work within a series of projects associated
with the European Materials Modelling Council (EMMC) aims at going beyond this
by achieving interoperability for all physical modelling and simulation methods,
including quantum mechanics, molecular modelling and simulation, and
continuum methods up to the macroscopic and process level.\citep{DeBaas17, LAL19}
Within this line of work, a Review of Materials Modelling (RoMM) was
conducted. \newydd{This review, which is now available
in its 6th edition,\citep{DeBaas17} resulted from work done within the European Commission
Directorate for Research and Innovation. Its long term goal is to increase
the competitiveness of European industries thanks to a stronger uptake
of materials modelling techniques for the different stages of manufacturing. 
Given the vast diversity of approaches and vocabularies used in subdomains of
the modelling world, RoMM proposes a harmonized language and a
classification of models to support communication across subdomains
and across roles (software developers, theoreticians, and industrialists).
A detailed explanation and discussion of this harmonized language,
containing numerous examples, is given
in the RoMM document.\citep{DeBaas17} On this basis,}
MODA (Model Data), a semi-intuitive graph language for
simulation workflows,\citep{SCLHGKWWGPAFCOGFBCW18} was introduced jointly with
a collection of further semantic
assets,\citep{Goldbeck19} including the European Materials \newydd{and}
Modelling Ontology (EMMO) which is under development by \citet{GFSG19}\ Compared to
generic workflow notations, MODA is tailored to optimally address aspects
that are specifically relevant to materials modelling, and it is based on
the RoMM terminology that was developed for the same purpose.
\newydd{Annex~II of the RoMM document\citep{DeBaas17} includes MODA
workflow description examples contributed by 36 projects from
the LEIT-NMBP line of the European Union's Horizon 2020
research and innovation programme.}

\newydd{The Virtual Materials Marketplace (VIMMP) is a platform,
presently under development, where services related to materials modelling
such as expertise, translation (from an industrial problem to a modelling
solution), software, model parameters, training, computing resources,
validation data, \etc, will be provided to end users. Accordingly, agents
on the market include individuals, groups, and institutions from the industrial and
academic world, such as modellers, software owners, and data providers.
By design, VIMMP is open to any interested provider, and the basis
for connecting their heterogeneous resources is a common language -- hence
the significance of a semantic approach.}
In this context, the present work
discusses the state of the art in semantic asset development
for simulation workflows in computational molecular engineering
and introduces a formalism based on ontologies
which can be employed to represent workflow metadata.
Thereby, it addresses the need for a formalized, machine-readable
representation of simulation workflows. This is done in a way
that facilitates an integration with the previous
and ongoing work done within the EMMC community, in particular
with MODA, which is the previous EMMC standard for
describing a simulation workflow.
To increase the expressive capacity and eliminate
ambiguities inherent in the semi-intuitive graph notation from MODA,
logical resources are introduced as entities related to the flow of information.
On the basis of \newydd{an improved}\marked{~\sout{this extended}} graph
notation \newydd{for simulation workflows}\marked{\sout{, referred
to as LDT graphs}} (\cf~Section \ref{subsec:ldt}),
an ontology for simulation, modelling, and optimization (OSMO)
is formulated (\cf~Section \ref{subsec:osmo})
which goes beyond MODA by being machine processable,
amenable to automated reasoning by semantic technology, and by which
workflow semantics in materials modelling are captured in a way that
is closely aligned and interoperable with the whole family
of semantic assets presently under development in the context
of the same infrastructures and projects.

To characterize software, in general, it is
possible to describe many different aspects
for a variety of purposes including, \eg, to identify, to
understand, to trade, or to use a given tool, and these descriptions
can be provided at multiple levels of detail.
Finding appropriate ways to cite software, recognize authorship, and give scientific credit to the
developers is a concern for different communities and key to making software development sustainable.
Along these lines, principles for software
citation have been proposed,\citep{SKN16, KBCHJC3DFG3H4JKKLRRTWZ19}
and the metadata schema CodeMeta\citep{CodeMeta} as well as
the citation file format CFF\citep{CFF} have been developed.
To describe simulation software
\marked{\sout{at the Virtual Materials Marketplace (VIMMP)} }\newydd{within VIMMP},
the VIMMP Software Ontology (VISO) is presented here (\cf~Section \ref{subsec:viso}), complementing OSMO.
The main focus of VISO is to facilitate the description
of software capabilities in computational molecular engineering.
%
%
Ontologies with a similar purpose, which describe the software from the
point of view of a scientist end user, have been developed in other fields,
\eg, the Software Ontology (SWO) in the area of
life sciences\citep{MBLIHPS14} and OntoSoft for geosciences;\citep{GRG15}
in particular, among other aspects, SWO also covers the implemented algorithms.
Here, with VISO, a comparable ontology is made available
for the area of materials modelling.
OSMO and VISO will be used by the VIMMP marketplace, its
components, and all interoperable platforms and environments, to represent
simulation workflows at a logical (\ie, non-technical) level and assist
in the selection of suitable software components and simulation platforms.\citep{Cavalcanti19}
\newydd{In particular, this work aims at facilitating the
description of information from model databases and parameterization
environments, such as Bottled SAFT\citep{EMM16, EJM17} or
the infrastructures designed by Hasse and collaborators,\citep{FBLHB18, SHVH19}
as well as workflow management systems (WMS)
and workflow repositories, \eg, TaLPas\citep{SVW19} and
exabyte,\cite{Bazhirov19} in a well-defined way to make such platforms
interoperable with VIMMP. Accordingly, the present work was
conducted as a collaboration of the TaLPas and VIMMP project consortia.}




The remainder of this article is structured as follows:
Section \ref{sec:semantic} discusses the challenge of achieving
interoperability of diverse tools and environments at the
levels of syntax, semantics, and pragmatics; VISO is introduced as a
formalism for simulation software descriptions.
In Section \ref{sec:workflow-environments}, workflow
management systems are discussed, and the environment developed
within the TaLPas project is presented; an example workflow is introduced,
concerning the parameterization of an equation of state (EOS)
by molecular simulation. This application scenario \newydd{is}
subsequently employed to illustrate the concepts from the present work.
Section \ref{sec:workflow-representation} comments on existing
formalisms by which simulation workflows can be represented at
a logical level, in particular the graph
notation from MODA. It is shown how
an \newydd{improved}\marked{~\sout{extended}} graph notation
can be employed to denote the flow of information
and dependencies between components of a workflow less
ambiguously, and the ontology OSMO is introduced, which provides
an additional layer of formalization to the characterization of
the involved classes of objects and the relations between them.
Finally, a conclusion is given in Section \ref{sec:conclusion}.

\section{Semantic interoperability}
\label{sec:semantic}

\subsection{Development of semantic assets}

Interoperability is the capacity of multiple codes or platforms, which
are not immediately compatible, to interact automatically by means of
a common representational system; \ie, whereas
for compatible environments, the sender needs to be familiar with the
concepts and data structures employed by the recipient, interoperability
does not require any bespoke tailoring to a specific target environment.
For a large number of (actual or potential) diverse interacting systems,
interoperability is the more scalable approach, since it does not
force the developers of each software or infrastructure to implement
all the formats required by a multitude of different codes. Instead, data
transferred between interoperable environments need to be
transformed to a single agreed intermediate stage by the sender,
and it is the duty of the recipient to implement the common
representational system adequately on his own side.

To facilitate interoperability, a common framework needs to be
established at three levels: Syntax, semantics, and pragmatics.\citep{AS10, WDBC16}
Thereby, syntactic interoperability refers to the standardization of
data formats and technical protocols for data transmission. However, beside
the need for a sender and the recipient to implement input/output
functionalities for the same format, they also need to agree on the
meaning of the communicated contents; this is semantic
interoperability. Only on this basis, full interoperability can be
achieved, which additionally requires an agreement on pragmatics, \ie,
the use of data,\citep{SDD06} including minimum standards for data and metadata curation,
research data management, validation, and assessment of data.
Pragmatic interoperability also concerns what to expect from an individual agent
with a particular social role,\citep{WDBBC16} \eg, a \textit{translator} who maps
a problem from industrial practice to viable solutions by computational
molecular engineering; significant efforts need to be devoted to
negotiating agreement on such expectations. Along these lines, in case of the
translator role, the EMMC has developed a pragmatic asset,
the Translators' Guide.\cite{HAKBMGHSSKKMID18}

Syntactic and semantic interoperability are closely related and
usually co-developed. If the focus is on file formats (hence, syntactic
interoperability leads the development), underlying
assumptions on the interpretation of the contents often remain
implicit; guidance on semantics is usually, if at all,
provided in human-readable form, \eg, in a user manual.
Obversely, if semantic interoperability leads the development,
standard serializations of data exist by which syntactic agreement
can be achieved in a straightforward way, such as the RDF/XML
format, the terse triple language (TTL), the hierarchical
data format HDF5, or the Allotrope data format.\citep{AH11, Schmitz16, OKSWC18}
The semantic assets usually take the form of metadata schemas
or ontologies, stating what classes of objects exist (in a certain domain,
\ie, the application field for which the schema or ontology is designed) and how
they can relate to each other.\citep{MWM08, AH11, MAD17, LAL19}
The approach based on semantic interoperability
has the advantage that the agreement on both the format and
the meaning is codified on the basis of definitions that can be processed
computationally, \eg, by automated logical reasoning.
In this way, the internal consistency of data sets can be checked, and
data from multiple sources can be integrated,\citep{MWT11} facilitating
more effective decision support systems.\citep{AB18}
Besides, the experience available so far suggests that
the development of ontologies can be a major step towards achieving
interoperability at all three levels, including pragmatics.\citep{DC07, Gan09, WDBC16}

As a prerequisite for such solutions, pre-existing dark data
need to be amended with appropriate metadata, in agreement with the established
semantic assets. This is a personnel-intensive task, for which
dedicated expertise is required, and which has to be repeated whenever
the semantic assets are replaced or undergo a major update.\citep{SD19}
Accordingly, it is important to reduce the risk that
significant changes become
necessary, which might disrupt backwards compatibility, at a
point when an ontology has already been employed to classify
great amounts of data and metadata.
Multiple perspectives, representative of the
envisioned community of future users, need to be involved
in the development of semantic assets from the first design onward.
Accordingly, requirements and experiences from
the VIMMP, TaLPas, and SmartNanoTox projects (\cf~Acknowledgment)
were taken into account for the present work, and ontology drafts
were made available to participants of the Horizon 2020 projects MarketPlace
and EMMC-CSA within the European Virtual Marketplace
Ontology working group.

\subsection{Software metadata at the Virtual Materials Marketplace \newydd{(VIMMP)}}
\label{subsec:viso}

The \newydd{ontology} \marked{\sout{VIMMP Software Ontology (}}VISO\marked{\sout{)}} was developed to support the
identification of suitable software tools and to standardize the
description of software tools as well as modelling and simulation approaches,
with the eventual aim of assisting users at accessing the VIMMP marketplace
infrastructure.
In particular, VISO will be used to structure the data ingest
about software tools at the VIMMP marketplace frontend. The same keywords will then be
available to the users to browse the tools and compare them.
Accordingly, the main purpose of VISO is to describe materials modelling software,
mostly addressing features and capabilities of models and solvers,
but also licensing, requirements (\eg, with respect to libraries and operating systems),
and compatibilities with other tools; a pre-release version of VISO (\texttt{viso-all-branches.ttl})
and an example of its use (\texttt{example-viso.ttl}) are included as Supporting Information.

The approach from RoMM, which is followed here, requires
a separation of the governing (\ie, constitutive) equations of a model into one or
multiple physical equations (PE) which pertain to the basic modelling approach and,
by definition, do not depend on the considered material, and
one or multiple materials relations (MR) which capture the
characteristics of the considered material.
Tab.~\ref{tab:viso-models} lists the main model types
considered by VISO. Therein, the PE type ID refers to a property
from OSMO, \cf~Section \ref{subsec:osmo},
where PEs that often occur within models
are classified into 25 categories on the basis of RoMM;
examples for this are provided as Supporting Information.
While the distinction between the PE and the MR may appear to be
straightforward from an abstract philosophical point of view,
its application to concrete models is often non-unique, and imperfect to a
certain extent, since the form and the content of a model cannot normally be
separated from each other completely.
Similarly, RoMM is also based on a strict
distinction between the model (\ie, the theoretical appproach) and
the solver (its numerical implementation);
accordingly, a \ttl{model\_feature} here
characterizes the underlying physical representation,
whereas a \ttl{sol\-ver\_feature} characterizes the implementation and
computational representation of the modelling approach by a numerical algorithm.
In practice, applying the split between model and solver features to a concrete
scenario poses similar challenges as in the case of the PE and the MR.
A prototypical example of this are thermostats: Depending on
the modelling perspective, they can either be seen as solver features
or, \eg, in dissipative particle dynamics (DPD), as fundamental
ingredients of the model. Moreover, in the latter case, there are
arguments both to include them in the PE, since they are necessary and their functional form is not
material dependent, or in the MR, since their parameters are related to the
material transport properties.
\newydd{The challenges mentioned above are unavoidable when logically
decomposing pre-existing complex models and software into a logical and simple structure.
Designed to combine information from a wide community of prospective
contributors and users, VISO provides a systematic approach for tackling
any ambiguities in this context.}

Below an upper level ({\tt viso-general}) that addresses general aspects
common to all software (\eg, the programming languages), we
split VISO into three branches, \ie, electronic (EL, {\tt viso-el}),
atomistic-mesoscopic (AM, {\tt viso-am}) for the two molecular
granularity levels from RoMM, and continuum (CO, {\tt viso-co}).
These branches expand on the model and solver features for each class.
The present formulation of these hierarchies was designed by evaluating a
representative set of software packages for CFD simulation as well as
quantum-mechanical density functional theory (DFT),
Monte Carlo (MC), MD, and DPD simulation.
Given that many model types can be described from several
points of view, VISO allows its users to represent certain
approaches in multiple ways; in such cases, the equivalence
relation \ttl{is\_modelling\_twin\_of} is employed to express that
despite being distinct in the ontology, 
certain instances of different classes can be employed as
representations of the same concepts.

\begin{table}[t]
   \caption{Models currently considered in developing \newydd{the VIMMP Software Ontology (}VISO\newydd{)}, associated \newydd{physical equation (}PE\newydd{)}, \newydd{materials relation (}MR\newydd{)}, and \newydd{physical equation type identifier (}PE
     type ID\newydd{); there, PE and MR are concepts from the Review of Materials Modelling\citep{DeBaas17} (RoMM), and the PE type ID is introduced in the present work}, \cf~Tab.\ \ref{tab:pe-types}.}
   \label{tab:viso-models}
   \small
   \begin{tabular}{llll}
   \hline
   Model type & Physical Equation (PE) & Materials
       Relation (MR) & PE type ID\\
   \hline
   DFT & Kohn-Sham
       eq. & Exchange-correlation  & EL.1 \\
     &  & functional, composition, \etc & \\
   MD  & Newton's II.\ law & Inter-particle
       potentials, & A.3, M.3  \\
 &  & composition, connectivity & \\
   MC & Partition function and ensemble- &
      Inter-particle potentials,  & A.4, M.4 \\   
 & average expressions & composition, connectivity & \\       
  DPD  & Newton's II.\ law (conservative  & Soft DPD + other potentials, & M.3 \\
 & force) + drag and random forces & composition, connectivity &\\
  CFD & Mass, momentum, and energy &  Constitutive relations (\eg,    & CO.2 \\       
 & transport eqs.\ (\eg, Navier-Stokes) & linear transport coefficients) \\
 EOS & Fundamental or thermal EOS & Functional form and & CO.5 \\
   & & parameters of the EOS \\
\hline
   \end{tabular}
\end{table}

Beside features, the other upper classes defined in VISO
are \ttl{software} (including operating systems, compilers, and software tools),
\ttl{agent}, \ttl{license}, \marked{\sout{and}} \ttl{programming\_language}\newydd{,
\ttl{model\-ling\_related\_entity} (including high level concepts related to modelling, such as model types),
\ttl{software\_interface} (based on the analogous class from SWO\cite{MBLIHPS14}; it includes,
\eg, graphical, command line, and application programming interfaces), and \ttl{soft\-ware\_update}.
The latter, in particular, allows to describe the addition/removal of features across versions of a tool.}

\begin{figure}[tb]
\includegraphics[width=15cm]{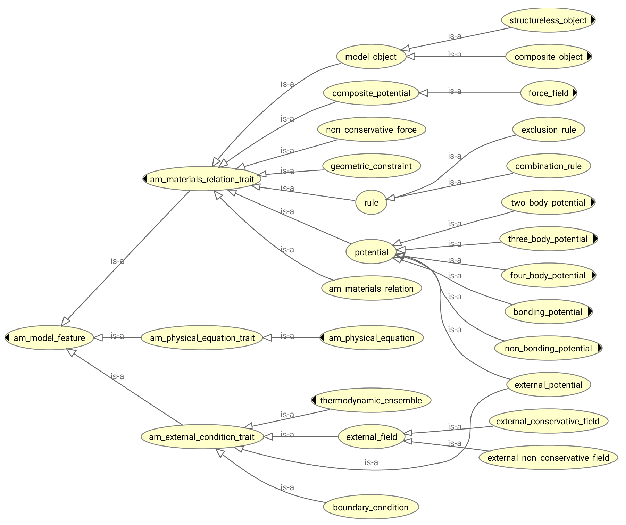}
\caption{VISO class \tttl{am\_model\_feature} and its subclasses. \newydd{This
    figure and similar ones in the present article have been generated using
    OWLViz.\citep{Horridge10}} \marked{\newydd{(Note: This figure was updated.)}}}
\label{fig:am_model_feature}
\end{figure}

The main relations defined in VISO to connect these components are briefly described in Tab.~\ref{tab:relations-in-viso},
and the direct subclasses of the \ttl{solver\_feature} class are listed in Tab.~\ref{tab:viso-solver-features}.
The \ttl{model\_feature} class has generally a richer structure, and we subdivide
it into the (non-disjoint) classes \ttl{physical\_equation\_trait}, \ttl{materials\_relation\_trait},
and \ttl{exter\-nal\_con\-dition\_trait}. As an example, Fig.~\ref{fig:am_model_feature} includes
the upper levels of the class hierarchy for the model features in particle-based models (\ie, in \texttt{viso-am}).
It can be seen that one of the categories for the MR traits
is \ttl{force\_field}, to be used for statements referring to popular
transferable group-contribution based methods (AUA,\citep{UBDBRF00} OPLS,\citep{Jorgensen86, JMT96} TraPPE,\citep{CPS01} \etc);
additionally, a finer level of description is available, explicitly identifying the functional forms
of the inter-particle potentials that are needed for the model of interest.
A possible use of VISO would be, given a force field or a set of MR traits, to identify a code that has them in its set of features.

\begin{table}
  \caption{Main relations, \ie, \tttl{owl:ObjectProperty} instances, defined in VISO,
   for which \tttl{software\_tool} is the domain (\ie, class of $X$). For more details,
   \cf~the Supporting Information. \newydd{By convention, the namespace \tttl{owl} is
   employed for keywords of the Web Ontology Language (OWL).}}
   \label{tab:relations-in-viso}
   \begin{tabular}{lll}
   \hline
      relation &  range & \textit{brief description}  \\
      (between $X$ and $Y$) & (\ie, class of $Y$) & \\
   \hline
      $X$ \tttl{has\_feature} $Y$
         & \tttl{model\_feature}
         & \textit{points to features of a tool} \\
     &  or \tttl{solver\_feature} & \\
      $X$ \tttl{is\_compatible\_with} $Y$  & \tttl{software\_tool}  &\textit{compatibility between tools}\\
   $X$ \tttl{is\_tool\_for\_model} $Y$    & \tttl{model\_type}  &  \textit{associates tools with models}\\
   $X$ \tttl{requires} $Y$  & \tttl{software} &  \textit{required \newydd{operating system or library}\marked{~\sout{OS and/or libraries}}}\\
   \hline
   \end{tabular}
\end{table}

\begin{table}[t]
  \caption{Classes of solver features defined within {\tt viso-el}, {\tt
      viso-am}, and {\tt viso-co}. The name\newydd{space} prefixes are shown in the upper row.
      \newydd{Therein, `el' represents the \textit{electronic} granularity level,
      `am' represents \textit{atomistic and mesoscopic}, and `co'
      stands for \textit{continuum}; these concepts are defined and discussed
      in the RoMM document.\citep{DeBaas17}} For further details, \cf~the Supporting Information.}
   \label{tab:viso-solver-features}
   \footnotesize
   \begin{tabular}{lll}
     \hline
    subclasses of \tttl{el\_solver\_feature}  & subclasses of \tttl{am\_solver\_feature} &
   subclasses of \tttl{co\_solver\_feature} \\
      (prefix: \tttl{viso-el})  & (prefix: \tttl{viso-am}) & (prefix: \tttl{viso-co}) \\
   \hline
   \tttl{basis\_set} & \tttl{barostat}&  \tttl{continuum\_mesh}\\ 
   \tttl{electron\_diagonalization} & \tttl{integrator}& \tttl{divergence\_scheme}\\
   \tttl{electron\_mixing} & \tttl{electrostatic\_solver} & \tttl{gradient\_scheme}\\
   \tttl{electron\_smearing} & \tttl{geometric\_constraint\_algorithm} & \tttl{spatial\_discretization\_scheme}\\
   \tttl{ionic\_relaxation} & \tttl{parallelization\_scheme} & \tttl{temporal\_discretization\_scheme}\\
   \tttl{kpoint\_mesh} & \tttl{sampling\_algorithm}  & \\
   \tttl{symmetry\_adapted\_solver} &  \tttl{thermostat}& \\
     \hline
   \end{tabular}
\end{table}

\section{Simulation workflows in materials modelling}
\label{sec:workflow-environments}

\subsection{Workflow management systems \newydd{(WMS)}}

There is a great variety of environments dealing with workflows.
A large number of workflow management systems (WMS)
has been implemented over the years, originating mainly from the fields of
data analysis and bioinformatics which in many cases
need to rely on large-scale automated computational pipelines.
The WMS are meant to facilitate an improved maintainability and robustness
compared, \eg, to plain shell scripts. For this purpose,
computations and data dependencies are linked logically, leaving details
of the task submission -- in many cases also including HPC load balancers -- to the WMS.
By abstracting from
all the logistics of file manipulation, copying
procedures, and data handling, the management systems thus allow researchers
to concentrate on improving the simulation or data-analysis workflow
instead of reimplementing standard procedures.\citep{LAS16}

Popular packages include Apache Airflow which allows
users to author workflows as directed acyclic graphs;\citep{Beauchemin19} in FireWorks,
workflows can be described in Python or markup languages and can
be monitored in web interfaces.\citep{JOCMQKBPRHGP15} Luigi,
pionieered by Spotify,
works on a similar basis and employs Python classes
for its workflow definition and task scheduling.\citep{EFFR17}
Snakemake, which is mainly aimed at bioinformatics, has its own domain-specific
language to define workflows, including many features oriented towards HPC;\citep{KR12}
beside, generic building environments like GNU make, which also underlies snakemake, can be used directly
to automate task dependence and workflow management for data analysis and simulation,
as in the case of the main component of the HOPS solver.\citep{RHLL02}
Moreover, several WMS,
including AiiDA,\citep{PCSMK16, MMCMGP17} Salome/YACS,\citep{RC07} and the
\newydd{present WMS for \textit{task-based load balancing and auto-tuning in particle-based
simulation}\citep{SVW19} (}TaLPas\newydd{)}\marked{~\sout{workflow and performance modelling environment}}, \cf~Section \ref{subsec:talpas}, have been designed particularly for simulation workflows in materials modelling.

\subsection{WMS \newydd{for task-based load balancing and auto-tuning in particle-based simulation (TaLPas)}\marked{~\sout{TaLPas workflow management system}}}
\label{subsec:talpas}

The TaLPas WMS was developed with the specific needs of the computational molecular engineering community in mind. Accordingly, it was designed to facilitate complex workflows, potentially consisting of a great number of individual simulation runs and data processing steps. Moreover, the molecular simulations performed within these workflows often need to be executed on HPC facilities due to their high computational demands, and the computational costs of single simulations (or tasks, in the case of task-based workflows) vary significantly depending on the simulation input parameters. Typical challenges hence include the management of a great amount of individual tasks, the organization of the results as well as the setup and execution of simulations on diverse and heterogeneous computer system environments and architectures.\citep{SVW19}

\begin{figure}[ht]
\includegraphics[width=14cm]{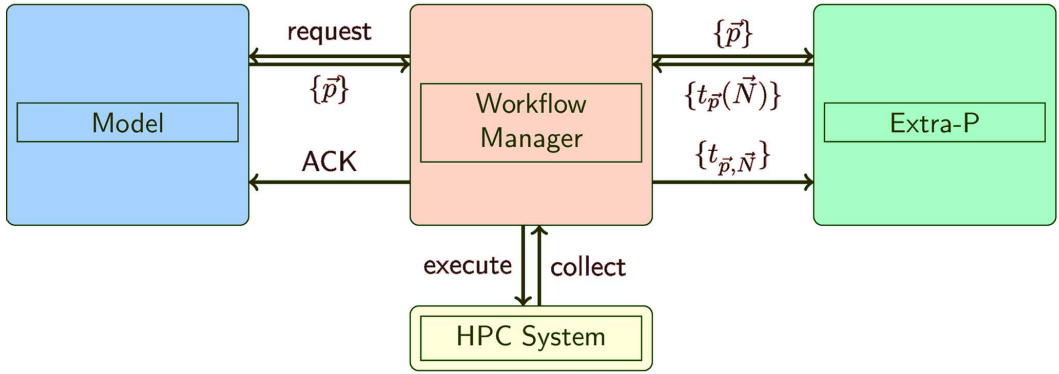}
\caption{Architecture of the workflow management system \newydd{(WMS) for \textit{task-based load balancing and auto-tuning in particle-based simulation} (TaLPas): The workflow manager requests tasks from the workflow model. The model responds with a set of parameters $\mathnewydd{\vec{p}}$ specifying the task to be executed. Tasks are then scheduled, and eventually, the results of the execution are collected by the workflow manager in combination with metadata on, e.g., the execution time and the error status. As soon as a task finishes, the model receives an acknowledgment (ACK) about the finished execution to decide on further execution steps. At the same time, performance-related information is communicated to the external performance model provider (here, Extra-P\citep{CBEHKSW16, SCHW17}). During the scheduling process, the workflow manager can query the performance provider for an estimate of the runtime $\mathnewydd{t_{\vec{p}}(\vec{N})}$ with a given amount of resources $\mathnewydd{\vec{N}}$. This performance model is used to improve the scheduling process.} \marked{\newydd{(Note: This figure was updated.)}}}
\label{fig:workflow-manager-design}
\end{figure}

The TaLPas WMS addresses these problems. Its overall architecture is shown in Fig.~\ref{fig:workflow-manager-design}. The main core of the environment is the definition of a workflow model. The model defines tasks, which are evaluated by the TaLPas workflow manager; a task includes information about the simulation parameters $\vec{p}$ as well as the simulation program and the commands required to execute it. The WMS comes with a set of selectable task schedulers which handle dependencies between tasks and facilitate their execution on a variety of different HPC systems. To access available computational resources on a HPC system, the scheduler uses a resource manager that keeps track of the availability and usage status of the provided nodes; it also processes the details of the execution for parallel applications via MPI \newydd{(message passing interface)}. For the determination of the task execution order, the WMS provides an interface allowing it to be extended by a performance efficiency provider such as Extra-P,\citep{CBEHKSW16, SCHW17} which estimates the performance and computational resource requirements using the simulation parameters specified in the task. Subsequently, the efficiency provider receives information about the actual time requirements $t_{\vec{p}, \vec{N}}$ of any completed tasks, by which Extra-P can refine its performance model\newydd{, cf.\ Fig.~\ref{fig:workflow-manager-design}}.

The TaLPas WMS handles data and files related to all tasks, automatically keeping them seperated by a configurable directory structure. Once the workflow has terminated, this makes it easy for the user to retrieve the simulation outcome. The WMS also collects additional information at runtime, which may help in the case that errors occur during the task execution.

The TaLPas WMS is immediately compatible with the molecular simulation codes \textit{ms2}, \cf~Rutkai \etal,\citep{RKGJSGBWSKRDHHV17} and \textit{ls1 mardyn}, \cf~Niethammer \etal\citep{NBBBEHWBGHVH14} Beyond case-by-case efforts at achieving compatiblity with individual software architectures, however, TaLPas aims at integrating a multitude of components for the development and optimization of complex task-based auto-tunable workflows. For this purpose, it is advantageous to achieve interoperability with the infrastructures developed on the basis of RoMM, MODA, and EMMO, and to describe simulation software and simulation workflows in terms of semantic assets formalized as ontologies.

\subsection{TaLPas WMS application scenario}
\label{subsec:eos-scenario}

EOS parameterization on the basis of high-throughput MC simulations was identified as a proof-of-concept application scenario for the development of the TaLPas WMS and its interoperability with other platforms, such as the VIMMP marketplace. To demonstrate the viability of the present approach, this is applied to phosgene (using the model by Huang \etal\citep{HHHV11}), building on previous work by Rutkai and Vrabec;\citep{RV15} there, the same problem was addressed without employing a dedicated WMS, and without characterizing the provenance of the EOS parameterization as well as the data obtained by molecular simulation.

\begin{figure}[ht]
\includegraphics[width=13.5cm]{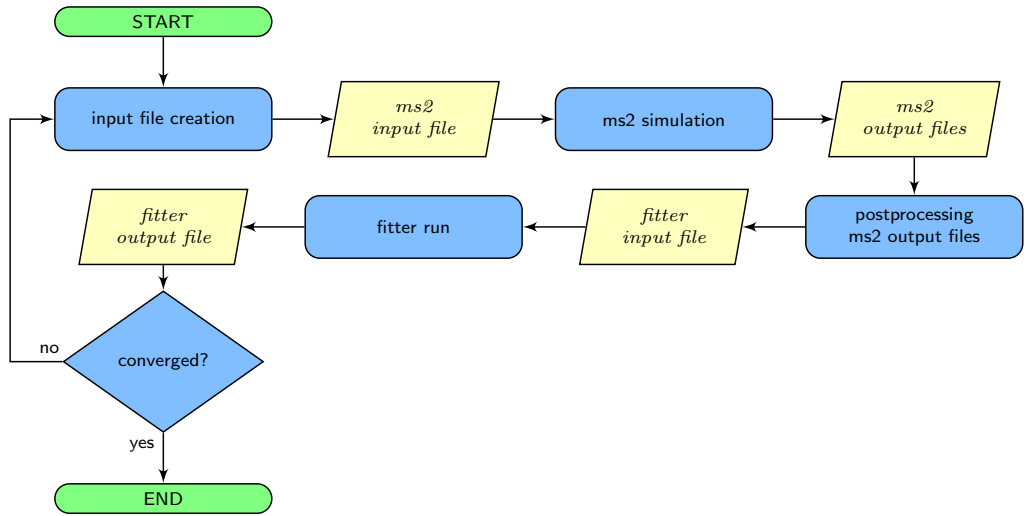}
\caption{Data flow and program execution of the \newydd{present equation of state (}EOS\newydd{) parameterization} workflow. Yellow boxes represent files, blue ones represent program executions, and diamonds represent conditions and branching.}
\label{fig:eos-dataflow}
\end{figure}

The present implementation addressing this class of problems uses sampling of state points and fitting with the method developed by \citet{SVW19}\ The corresponding workflow can be implemented using the \textit{ms2} simulation program. The data flow and steps to be performed are depicted in Fig.~\ref{fig:eos-dataflow}, with a focus on technical input/output using files; \cf~Section \ref{sec:workflow-representation} and the Supporting Information for a representation at the logical level, abstracting from the technical implementation of data transfer. A set of thermodynamic states, each of which is defined by the density and the temperature, is simulated in the canonical ensemble. The output of the simulations is processed to obtain multiple derivatives of the Massieu potential, following the formalism proposed by Lustig.\citep{Lustig11, Lustig12} The Massieu potential derivatives are used to generate the input for an EOS fitter. The result of the fit is not very accurate at the beginning. To increase the accuracy, additional state points are simulated in a series of iterations. The choice of the state points has a considerable influence on the convergence behaviour; in particular, state points close to the vapour-liquid coexistence curve are good candidates to consider for additional simulations. Therefore, intermediate evaluations are performed to refine the state points in an efficient way.

\newydd{A corresponding workflow model was created for the TaLPas WMS. The workflow model implements the steps from Fig.~\ref{fig:eos-dataflow} as well as the application programming interface of the TaLPas WMS. The model is handed over to the WMS to be executed. The structure and the most important parts of the workflow model are outlined below:}

\begin{verbatim}
class Model:
   def name(self)
      """Returns the name of the workflow"""
   def __init__(self)
      """Initializes the workflow parameters for the
~     refinement as well as an initial set of state
~     points to be processed."""
   def get_task(self)
       """Returns task objects which are intended to
~      be executed on the HPC resources. Ends the
~      workflow by returning an final task object
~      when convergence is achieved."""
   def deploy(self, task, np, mpi)
       """Generates all necessary ms2 input files and
~      constructs the final MPI command to execute ms2
~      on the HPC system."""
   def record_result(self, task)
      """Records the result of a ms2 simulation
~     run."""
   def createEosInputFromResults(self)
      """Implements the post processing step
~     converting the ms2 output files into an EOS
~     fitter input file"""
   def fitVleCurve(self)
      """Executes the EOS fitter with the generated
~     EOS input file"""
   def refine_around_critical_point(self)
      """Performs refinement around the current
~     critical point creating new state point to be
~     evaluated."""
   def refine_around_VLE(self)
      """Performs refinement around the VLE curve
~     creating new state point to be evaluated."""
\end{verbatim}

\newydd{The method \kbd{get\_task()} returns the task object, which is prepared for execution on the available HPC resources by the TaLPas WMS. Thereby, a task object is communicated in JavaScript Object Notation (JSON). A typical example for a task is given below:} 

\begin{verbatim}
{
  "ID": 53,
  "params": {
    "T": 1.5,
    "rho": 0.01,
    "step": 0
  },
  "taskdir": "workflow/results/T_1.5/rho_0.01/step_0",
  "deploy": {
    "NP": 4,
    "cmd": ["mpirun", "-np", "4", "./ms2",
~   "EOS_phosgene.par"],
    "nodes": [...]
  },
  "env": "...",
  "starttime": "2019-08-13T15:49:37.938883",
  "endtime": "...",
  "returncode": ...
}
\end{verbatim}

\newydd{The method \kbd{deploy()} creates all necessary input for the execution of \textit{ms2} as well as the final MPI command and stores the execution information in a task object. The method \kbd{get\_task()} hands those task objects over to the WMS for execution. The WMS checks for available resources and starts \textit{ms2} using MPI according to the task object. As soon as the task finishes, the method \kbd{record\_result()} is called so that the workflow model can record the result for the fitting and iterative refinement step:}

\section{Representation of simulation workflows}
\label{sec:workflow-representation}

\subsection{Graph and diagram notation approaches}
\label{subsec:ldt}

The present section discusses how the simulation workflow graphs from
MODA can be \newydd{improved}\marked{~\sout{extended}} to
account for logical data transfer (LDT)
and dependencies between workflow elements in a more explicit way;
on this basis, in Section \ref{subsec:osmo}, OSMO is introduced as
an ontology that formalizes the relations visualized by the
LDT graph notation and is closely aligned with MODA in
its description of the elementary parts of the workflow.

For simulation workflows
(and workflows more generally), highly developed
formal descriptions exist, including ontologies and graph
languages.\citep{PCH05, RGS14, KTC17}
Diagram-like notations, which in most cases can be represented
as graphs -- in the sense employed in graph theory, \ie,
as structures that consist of a) nodes and b) edges that connect the nodes -- or
similar structures such as hypergraphs,\cite{Comuzzi19, WWJ19} exist at
various degrees of elaboration. At an informal level, this
may include, \eg, intuitive sketches drawn on a board to assist a
discussion, whereas a great degree of standardization and
formalization can yield highly elaborate systems such as
machine-readable representations of process flow
diagrams.\marked{~\sout{Semantically driven developments aim at
interoperability, \eg, by ontologies with relations and
objects}} \newydd{Ontologies and relations between objects} can
be visualized as graphs;\cite{MWM08}
syntactically, graph languages can be defined by graph grammars\citep{EEKR99}
or other formal approaches such as type graphs.\citep{CKN19}
In particular, such approaches have been applied to specify
and visualize concurrent and distributed
algorithms \newydd{and workflows}.\citep{BMMS01, SCHW17, CGCG18}
Often, however, semi-formal specifications of
diagram-like notations are provided, which
are not machine-processable, but intelligible to human users
and standardized to an intermediate extent.

\begin{figure}
  \includegraphics[width=15cm]{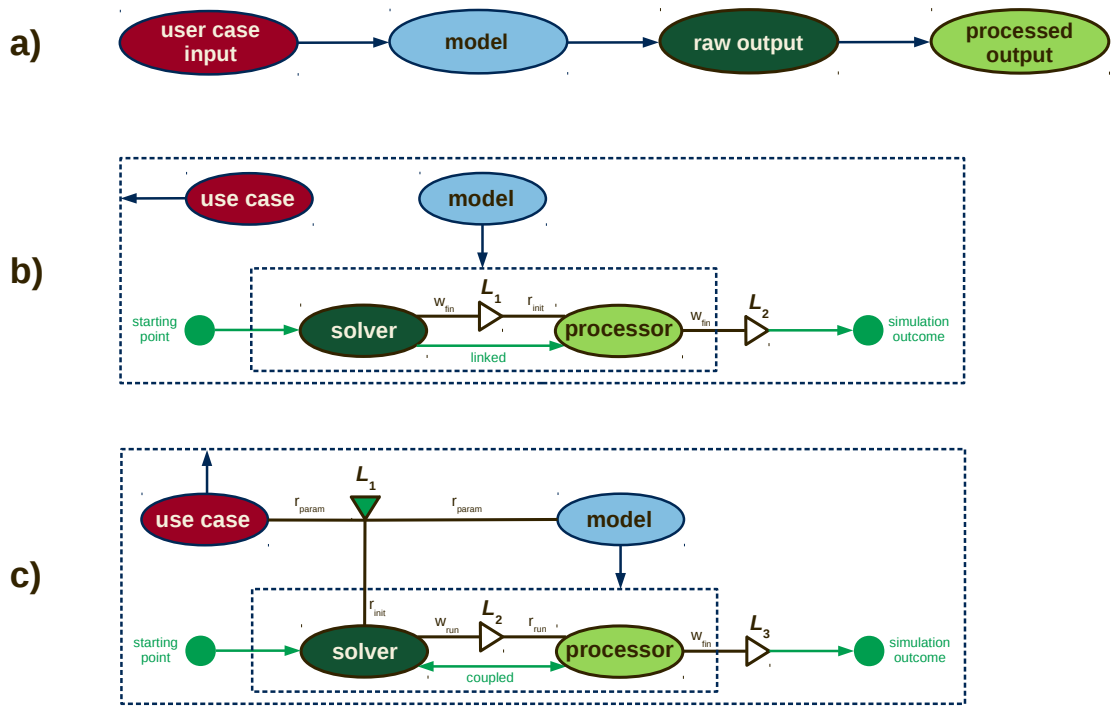}
  \caption{Comparison between the \newydd{Model Data\citep{SCLHGKWWGPAFCOGFBCW18} (}MODA\newydd{)}
  and \newydd{logical data transfer (}LDT\newydd{)} workflow graph
  notations: a) MODA graph where input characterizing a use
  case, a model, the raw output of a solver, and the
  processed output of a postprocessor are connected by blue
  arrows. b), c) Two LDT graphs corresponding to different
  scenarios which, in MODA, would both be represented by a).}
  \label{fig:ambiguity}
\end{figure}

The level of formalization of MODA,
a core building block of the EMMC approach to interoperability
in materials modelling, is at an intermediate stage:
It is defined by a CEN Workshop Agreement\citep{SCLHGKWWGPAFCOGFBCW18} \newydd{(CWA)
of the European Committee for Standardization (CEN)},
and \marked{\sout{the most recent revisions}}\newydd{Annex II} of RoMM
include\newydd{s} a catalogue of MODA examples.\citep{DeBaas17}
However, the descriptors for use cases, models, solvers,
and processors in MODA are restricted to plain-text entries,
which cannot be easily integrated with other elements of
the EMMC-governed semantic technology framework. Moreover, the
semantics of the characteristic blue-arrow edges that
connect the \textit{sections} (\ie, nodes) of a MODA workflow graph
are not defined by the CWA\marked{~\sout{CEN Workshop Agreement}};
arrows can represent any association between elements.
This is illustrated here
by a simple MODA graph consisting of four
sections, \cf~Fig.\ \ref{fig:ambiguity}a); \nb\ that in
MODA this graph would supplemented by a structured
plain-text description of the associated use case, model,
solver, and processor entities. However, the semantics of the blue
arrows is subject to the interpretation by a human reader.

Therefore, MODA is not sufficiently
unambiguous at the level of the workflow graph notation;
moreover, it is not an ontology, which would
be needed to combine it with the EMMO, other ontologies,
and semantic-technology driven infrastructures.
On the other hand, existing approaches from the literature
cannot be mapped to MODA in a straightforward way;
this also holds for OntoCAPE,\cite{MWM08}
the ontology that was developed to support CAPE-OPEN.\citep{BP02, LCZ09}
It is hence a necessity to develop a more elaborate graph notation
and an ontology on the basis of MODA.

The LDT notation clarifies how the use case, model,
solver, and processor entities in a workflow relate to each other,
\cf~Fig.\ \ref{fig:ambiguity}b) and c).
Therein, ellipses represent sections (\ie, use cases, models, solvers,
and processors); green circles and green arrows represent
coupling and linking of elements \newydd{(as per RoMM\cite{DeBaas17})},
dependencies concerning the order of execution,
and aspects related to concurrency and synchronization.
Blue arrows point from use cases and models to the part of
the workflow to which these entities apply; in particular, if
a model applies to a part of a workflow that contains solver
entities, these solvers are numerical implementations of this model.
Triangles are \textit{logical resources} which are employed to
describe how information is transferred between the sections.
The triangles point from the source of data
to the destination of data. If a triangle is filled (green colour),
this implies that a user interaction can occur concerning the
data stored at the respective logical resource;
this interactivity can consist of any potential steering
or input by a user at workflow initialization, execution,
or finalization time.

\begin{figure}
  \includegraphics[width=17cm]{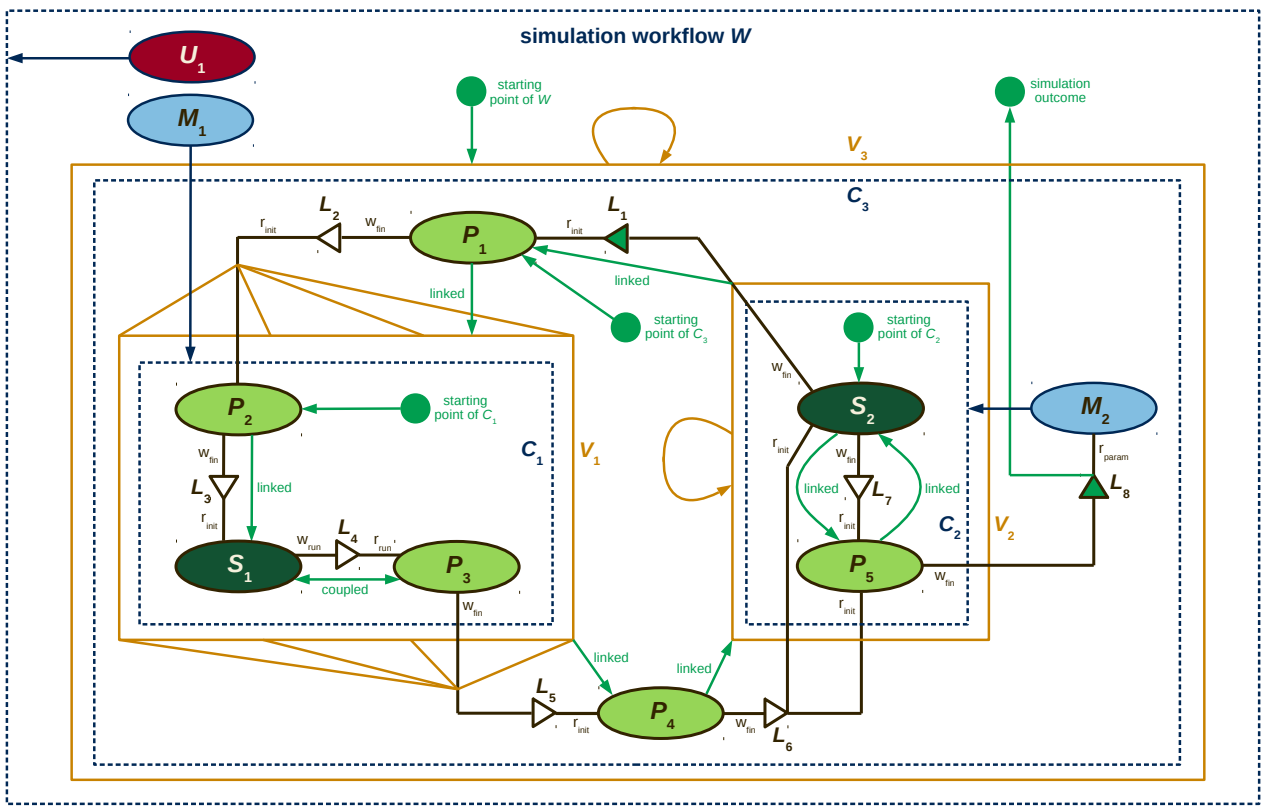}
  \caption{LDT graph representation of the example scenario
  from Section~\ref{subsec:eos-scenario}, where simulations
  on the basis of an intermolecular pair potential (model $M_1$,
  implemented by the solver $S_1$) are conducted to parameterize
  an EOS (model $M_2$, implemented by the
  solver $S_2$) for the purpose of predicting
  the thermodynamic behaviour of phosgene (use case $U_1$).
  The golden solid box with four golden lines at the entry
  and exit points (virtual graph $V_1$) represents a
  concurrent execution of multiple instances of the included
  blue dashed box (concrete graph $C_1$),
  and the golden solid boxes with golden loop-like arrows (virtual
  graphs $V_2$ and $V_3$) represent iterative
  executions of the included blue dashed boxes (concrete
  graphs $C_2$ and $C_3$).
  A characterization of this workflow following
  \newydd{the ontology for simulation, modelling, and optimization (}OSMO\newydd{)},
  in \newydd{terse triple language (}TTL\newydd{)} format, is included as Supporting Information
  (\texttt{eos-parameterization.ttl});
  see also Tab.~\ref{tab:ldt-in-osmo} for the relations from OSMO
  corresponding to the visual features from LDT graphs.}
  \label{fig:example-workflow}
\end{figure}

In this way, different workflows, which in MODA would
be ambiguously represented by the same graph, \eg, by
Fig.\ \ref{fig:ambiguity}a), can be distinguished:
\begin{itemize}
   \item{} In the case of the workflow represented by the LDT
      graph from Fig.\ \ref{fig:ambiguity}b), the model \textit{applies
      to} (blue arrow), \ie, is solved and taken into account by, a solver and a
      processor. The use case \textit{applies to} the entire workflow.
      The \textit{starting point} (green bullet) of the workflow is the solver,
      which \textit{is linked to} (green arrow) the processor.
      Linking here refers to a sequential dependency, \ie, the solver needs
      to terminate for the processor to start; therefore, in this case,
      the processor is a \textit{postprocessor}.
      Upon termination, the solver \textit{writes finally} (w$_\mathrm{fin}$)
      information to the logical resource labelled $L_1$, which is
      \textit{read initially} (r$_\mathrm{init}$) by the processor;
      \nb, writing and reading here represents any mechanism of
      dealing with information, irrespective of the way
      in which this is implemented. Eventually, the
      results computed by the postprocessor and written to the logical resource $L_2$,
      constitute the overall \textit{simulation outcome} (green bullet).

   \item{} The workflow from Fig.\ \ref{fig:ambiguity}c) deviates from
      this in ways that would be hard or impossible to make explicit in
      MODA notation. Here, the solver \textit{is coupled with} the processor
      (bidirectional green arrow), \ie, the execution of the two sections
      is synchronized. Accordingly, in this case, the processor is a \textit{coupled
      processor} instead of a postprocessor. Moreover, the use case and
      the model are parameterized, \ie, they \textit{read
      parameters} (r$_\mathrm{param}$) from a logical resource (here $L_1$)
      that is \textit{interactive} (green triangle).
      Input from $L_1$ is also used by the solver upon
      initialization. 
\end{itemize}
An LDT graph for the EOS parameterization example
scenario from Section~\ref{subsec:eos-scenario} is shown in
Fig.~\ref{fig:example-workflow}; see also the internal representation
from the TaLPas workflow environment, \cf~Fig.~\ref{fig:eos-dataflow}.
As in the case of a MODA graph, a description of the
use case, model, solver, and processor entities
entities always needs to be provided additionally,
which can be done at the ontological level following OSMO as
outlined in Section \ref{subsec:osmo} and Tab.\ \ref{tab:use-case-aspects}.

\subsection{Ontology for simulation, modelling and optimization \newydd{(OSMO)}}
\label{subsec:osmo}

\subsubsection{OSMO, the ontology version of \newydd{the Model Data (}MODA\newydd{) standard}}

\begin{figure}
   \includegraphics[width=17cm]{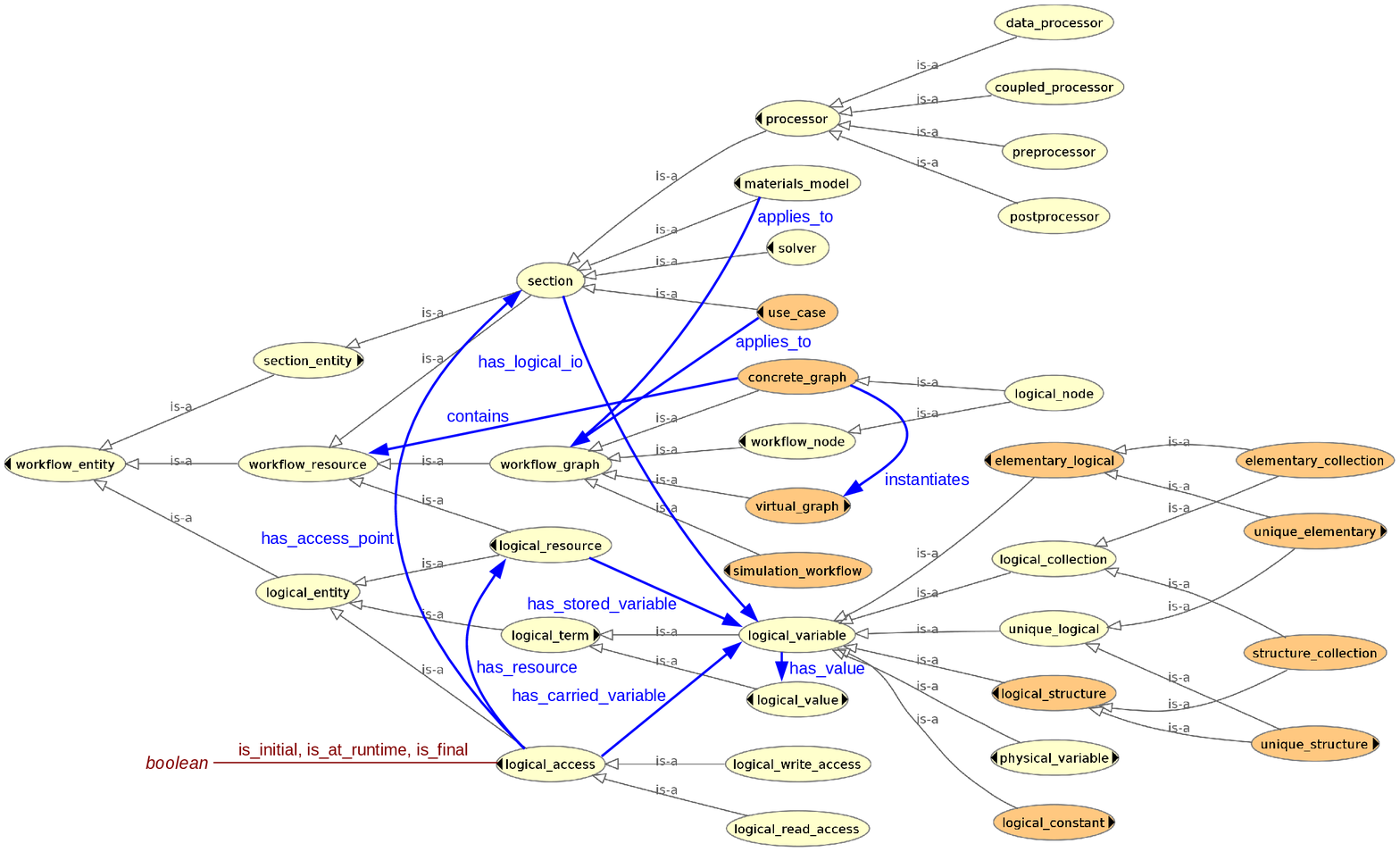}
   \caption{Workflow-related part of the OSMO class diagram, including
   the \tttl{rdfs:subClassOf} relation between classes (grey arrows) and
   selected additional relations defined in OSMO (blue arrows),
   as well as boolean features, \ie, instances of \tttl{owl:DatatypeProperty},
   defined for the class \tttl{logical\_access} (dark red).
   \newydd{By convention, the namespace \tttl{rdfs} is
   employed for RDF Schema (RDFS) keywords.} 
   \marked{~\newydd{(Note: This figure was updated.)}}}
  \label{fig:workflow-subclass-diagram}
\end{figure}

The ontology \marked{\sout{for simulation, modelling, and optimization (}}OSMO\marked{\sout{)}}, which
is introduced here, is based on the vocabulary
and the approach from RoMM;\citep{DeBaas17}
its representation of use cases, solvers, models, and processing
is directly based on MODA,\citep{SMKBKSGRRKLGLVH17} and the representation of workflows
is based on the LDT notation, \cf~Section \ref{subsec:ldt},
which is itself also an extension of MODA.
The class hierarchy for the part of OSMO related to simulation workflows
is shown in Fig.\ \ref{fig:workflow-subclass-diagram}, including some of the
relations that correspond to the visual features of the LDT graph notation;
these relations are summarized in Tab.\ \ref{tab:ldt-in-osmo}. By providing a
common semantic basis for workflows that were designed with different
tools, OSMO can be employed to consistently integrate data provenance
descriptions for materials modelling data from diverse sources.

\begin{table}
   \caption{Relations, \ie, \tttl{owl:ObjectProperty} instances, defined in OSMO to represent features of simulation workflows, with the corresponding symbols in the LDT graph notation; for a complete specification, \cf~the Supporting Information.}
   \label{tab:ldt-in-osmo}
   \scriptsize
   \begin{tabular}{llll}
   \hline
      relation & domain & range & LDT symbol \\
      (between $X$ and $Y$) & (\ie, class of $X$) & (\ie, class of $Y$) & \textit{italics: concise explanation} \\
   \hline
      $X$ \tttl{applies\_to} $Y$
         & \tttl{use\_case}
         & \tttl{workflow\_graph}
         & blue arrow from ellipse $X$ to box $Y$ \\
         & or \tttl{\newydd{materials\_model}} & & \textit{$Y$ deals with $X$} \\
      $X$ \tttl{contains} $Y$ 
         & \tttl{\newydd{concrete\_graph}}
         & \tttl{workflow\_resource}
         & $Y$ is graphically located inside $X$ \\
         & & & \textit{$Y$ occurs within $X$} \\
      $X$ \tttl{has\_access\_point} $Y$
         & \tttl{logical\_access}
         & \tttl{section}
         & $X$ is a line connected to ellipse $Y$ \\
         & & & \textit{LDT by $X$ involves section $Y$} \\
      $X$ \tttl{has\_carried\_variable} $Y$
         & \tttl{logical\_access}
         & \tttl{logical\_variable}
         & not visualized \\
         & & & \textit{LDT by $X$ concerns a transfer of $Y$} \\
      $\mathnewydd{X}$ \tttl{\newydd{has\_internal\_lv}} $\mathnewydd{Y}$
         & \tttl{\newydd{section}}
         & \tttl{\newydd{logical\_variable}}
         & \newydd{not visualized} \\
         & & & \textit{$\mathnewydd{Y}$ \newydd{is a logical variable that occurs in} $\mathnewydd{X}$} \\
      $\mathnewydd{X}$ \tttl{\newydd{has\_logical\_io}} $\mathnewydd{Y}$
         & \tttl{\newydd{section}}
         & \tttl{\newydd{logical\_variable}}
         & \newydd{not visualized} \\
         & & & \textit{$\mathnewydd{X}$ \newydd{reads or writes} $\mathnewydd{Y}$} \\
      $X$ \tttl{has\_resource} $Y$
         & \tttl{logical\_access}
         & \tttl{logical\_resource}
         & $X$ is a line connected to triangle $Y$ \\
         & & & \textit{LDT by $X$ involves resource $Y$} \\
      $X$ \tttl{has\_simulation\_outcome} $Y$
         & \tttl{simulation\_workflow}
         & \tttl{logical\_node}
         & arrow from $Y$ to a green bullet \\
         & & & \textit{resource at $Y$ contains end result of $X$} \\
      $X$ \tttl{has\_starting\_point} $Y$
         & \tttl{workflow\_graph}
         & \tttl{workflow\_node}
         & green bullet with an arrow to $Y$ \\
         & & & \textit{(sub-)workflow $X$ begins at position $Y$} \\
      $X$ \tttl{has\_stored\_variable} $Y$
         & \tttl{logical\_resource}
         & \tttl{logical\_variable}
         & not visualized \\
         & & & \textit{$Y$ can be read from or written to $X$} \\
      $X$ \tttl{has\_terminal\_point} $Y$
         & \tttl{workflow\_graph}
         & \tttl{workflow\_node}
         & not visualized \\
         & & & \textit{(sub-)workflow $X$ ends at position $Y$} \\
      $\mathnewydd{X}$ \tttl{\newydd{has\_value}} $\mathnewydd{Y}$
         & \tttl{\newydd{logical\_variable}}
         & \tttl{\newydd{logical\_value}}
         & \newydd{not visualized} \\
         & & & \textit{$\mathnewydd{X}$ \newydd{has the value} $\mathnewydd{Y}$} \\
      $X$ \tttl{instantiates} $Y$ 
         & \tttl{concrete\_graph}
         & \tttl{virtual\_graph}
         & golden solid box around blue dashed box \\
         & & & \textit{$Y$ is conditional/multiple execution of $X$} \\
      $X$ \tttl{is\_coupled\_with} $Y$
         & \tttl{workflow\_graph}
         & \tttl{workflow\_graph} 
         & bidirectional green arrow \\
         & & & \textit{$X$ and $Y$ are coupled, \ie, synchronized} \\
      $X$ \tttl{is\_direct\_cause\_of} $Y$
         & \tttl{workflow\_graph}
         & \tttl{workflow\_graph}
         & green arrow from $X$ to $Y$ \\
         & & & \textit{$X$ needs to terminate before $Y$ can begin} \\
      $X$ \tttl{is\_linked\_to} $Y$
         & \tttl{workflow\_graph}
         & \tttl{workflow\_graph}
         & green arrow from $X$ to $Y$ or vice versa \\
         & & & ($X$ \tttl{is\_direct\_cause\_of} $Y$ \\
         & & & or $Y$ \tttl{is\_direct\_cause\_of} $X$) \\
   \hline
   \end{tabular}
\end{table}

\begin{figure}
  \includegraphics[width=16.25cm]{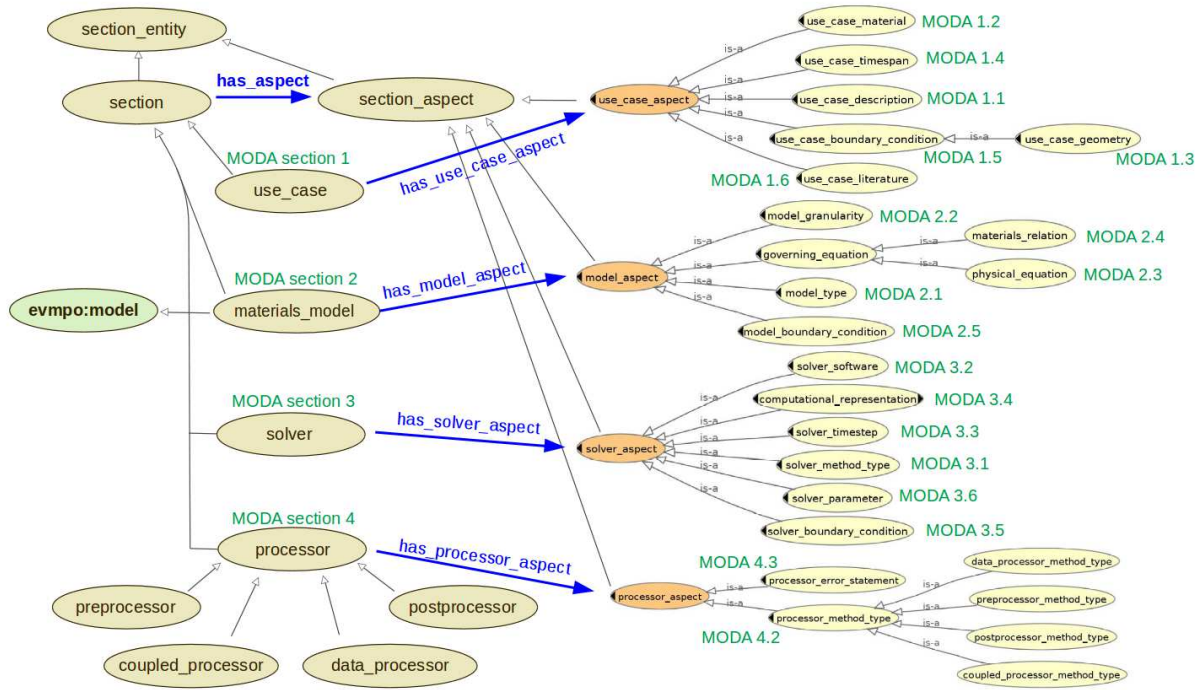}
  \caption{OSMO class \tttl{section\_entity} and its subclasses. The subclass relation is represented by grey arrows, and \tttl{has\_aspect} as well as its major subproperties are represented by blue arrows; entry numbers from MODA corresponding to the OSMO aspects, \cf~Tabs.~\ref{tab:use-case-aspects} -- \ref{tab:processor-aspects}, are denoted in green colour. The class \tttl{evmpo:model} represents a concept from the European Virtual Marketplace Ontology (work in progress), which defines a model by equivalence to the same concept from the \newydd{European Materials and Modelling Ontology\cite{GFSG19} (}EMMO\newydd{)}; as a special type of models, \tttl{materials\_model} from OSMO is a subclass of \tttl{evmpo:model}. For further details, \cf~the Supporting Information.}
  \label{fig:osmo-aspects}
\end{figure}

The detailed description of the four types of section
entities (use cases, models, solvers, and processors) in OSMO
follows the specification from MODA closely,
\cf~Tabs.~\ref{tab:use-case-aspects} -- \ref{tab:processor-aspects} for
the list of aspects (\ie, section descriptors)
and Fig.~\ref{fig:osmo-aspects} as well as
the Supporting Information for technical details.

\begin{table}
   \caption{Aspects of a \tttl{use\_case}, with the corresponding MODA entry numbers\citep{SCLHGKWWGPAFCOGFBCW18}}
   \label{tab:use-case-aspects}
   \footnotesize
   \begin{tabular}{lll}
   \hline
      OSMO aspect class name ~&~ MODA ~&~ aspect and content description (see TTL for details) \\
   \hline
      \tttl{use\_case\_description} ~&~ 1.1 ~&~ \textit{use case summary intended for human readers} \\
         ~&~ ~&~ content: plain text (elementary datatype \tttl{string}) \\
      \tttl{use\_case\_material} ~&~ 1.2 ~&~ \textit{characterization of the considered material} \\
         ~&~ ~&~ content: OSMO/EMMO\citep{GFSG19} class \tttl{material} \\
      \tttl{use\_case\_geometry} ~&~ 1.3 ~&~ \textit{description of the geometry of the considered system} \\
         ~&~ ~&~ content: plain text, OSMO class \tttl{condition} \\
      \tttl{use\_case\_timespan} ~&~ 1.4 ~&~ \textit{time interval of a process considered in the use case} \\
         ~&~ ~&~ content: OSMO class \tttl{timespan\_information} \\
      \tttl{use\_case\_boundary\_condition} ~&~ 1.5 ~&~ \textit{thermodynamic, spatio-temporal, or other condition} \\
         ~&~ ~&~ content: plain text, OSMO class \tttl{condition} \\
      \tttl{use\_case\_literature} ~&~ 1.6 ~&~ \textit{literature reference related to the use case} \\
         ~&~ ~&~ content: OTRAS/IAO\citep{CS15} class \tttl{citation} \\
   \hline
   \end{tabular}
\end{table}

\begin{table}
   \caption{Aspects of a \tttl{materials\_model}, with the corresponding MODA entry numbers\citep{SCLHGKWWGPAFCOGFBCW18}}
   \label{tab:model-aspects}
   \footnotesize
   \begin{tabular}{lll}
   \hline
      OSMO aspect class name ~&~ MODA ~&~ aspect and content description (see TTL for details) \\
   \hline
      \tttl{model\_type} ~&~ 2.1 ~&~ \textit{PE type following RoMM\citep{DeBaas17} and Section \ref{subsubsec:physical-equations}} \\
         ~&~ ~&~ content: OSMO class \tttl{physical\_equation\_type} \\
      \tttl{model\_granularity} ~&~ 2.2 ~&~ \textit{granularity level following RoMM\citep{DeBaas17} and Section \ref{subsubsec:physical-equations}} \\
         ~&~ ~&~ content: \tttl{ELECTRONIC}, \tttl{ATOMISTIC}, \tttl{MESOSCOPIC}, or \tttl{CONTINUUM} \\
      \tttl{physical\_equation} ~&~ 2.3 ~&~ \textit{detailed description of the employed PE} \\
         ~&~ ~&~ content: plain text (\ie, \tttl{string}), OSMO class \tttl{condition} \\
      \tttl{materials\_relation} ~&~ 2.4 ~&~ \textit{MR following RoMM\citep{DeBaas17}} (\eg, a pair potential) \\
         ~&~ ~&~ content: plain text, OSMO class \tttl{condition} \\
      \tttl{model\_boundary\_condition} ~&~ 2.5 ~&~ \textit{statement on boundary conditions applied to the model} \\
         ~&~ ~&~ content: plain text, OSMO class \tttl{condition} \\
   \hline
   \end{tabular}
\end{table}

\begin{table}
   \caption{Aspects of a \tttl{solver}, with the corresponding MODA entry numbers\citep{SCLHGKWWGPAFCOGFBCW18}}
   \label{tab:solver-aspects}
   \footnotesize
  \begin{tabular}{lll}
   \hline
      OSMO aspect class name ~&~ MODA ~&~ aspect and content description (see TTL for details) \\
   \hline
      \tttl{solver\_method\_type} ~&~ 3.1 ~&~ \textit{description of the numerical approach} (\eg, MD) \\
         ~&~ ~&~ content: plain text (\ie, \tttl{string}), VISO class \tttl{solver\_feature} \\
      \tttl{solver\_software} ~&~ 3.2 ~&~ \textit{employed software that implements the approach} \\
         ~&~ ~&~ content: plain text, VISO class \tttl{software\_tool} \\
      \tttl{solver\_timestep} ~&~ 3.3 ~&~ \textit{numerical time step employed by the solver} (if applicable) \\
         ~&~ ~&~ content: plain text, time expressed following QUDT/EMMO\citep{GFSG19, HKHS19} \\
      \tttl{computational\_representation} ~&~ 3.4 ~&~ \textit{describes how the solver represents the governing equations} \\
         ~&~ ~&~ content: plain text, OSMO class \tttl{condition} \\
      \tttl{solver\_boundary\_condition} ~&~ 3.5 ~&~ \textit{numerical boundary conditions applied within the solver} \\
         ~&~ ~&~ content: plain text, OSMO class \tttl{condition} \\
      \tttl{solver\_parameter} ~&~ 3.6 ~&~ \textit{parameter of the solver} \\
         ~&~ ~&~ content: OSMO class \tttl{\newydd{logical\_variable}} \\
   \hline
   \end{tabular}
\end{table}

\begin{table}
   \caption{Aspects of a OSMO \tttl{processor}, with the corresponding MODA entry numbers\citep{SCLHGKWWGPAFCOGFBCW18}}
   \label{tab:processor-aspects}
   \footnotesize
  \begin{tabular}{lll}
   \hline
      OSMO aspect class name ~&~ MODA ~&~ aspect and content description (see TTL for details) \\
   \hline
      \tttl{processor\_method\_type} ~&~ 4.2 ~&~ \textit{describes the methodology employed by the processor} \\
         ~&~ ~&~ content: plain text (\ie, \tttl{string}) \\
      \tttl{processor\_error\_statement} ~&~ 4.3 ~&~ \textit{uncertainty, error, or deviation from the most accurate value} \\
         ~&~ ~&~ content: \newydd{plain text, VIVO\citep{Cavalcanti19} class \tttl{accuracy\_assertion}} \\
   \hline
   \end{tabular}
\end{table}

\newydd{For example, metadynamics and its variants are solver algorithms; cf. RoMM,\cite{DeBaas17} p.~59 (``accelerated methods in molecular dynamics''). In MODA,\citep{SCLHGKWWGPAFCOGFBCW18} this is specified by entry 3.1. In OSMO, the MODA entry 3.1 corresponds to the aspect \tttl{osmo:solver\_method\_type}, cf.~Tab.~\ref{tab:solver-aspects}, which points to a \tttl{viso:solver\_feature} object. For this purpose, the atomistic-mesoscopic branch of VISO provides the class \tttl{viso-am:sampling\_algorithm}, cf.~Tab.~\ref{tab:viso-solver-features}, which is a subclass of \tttl{viso:solver\_feature}. Accordingly, the fact that a solver employs well-tempered metadynamics can be denoted as follows:}

\begin{verbatim}
:SX a osmo:solver;
   osmo:has_solver_method_type [
      a osmo:solver_method_type;
      osmo:has_aspect_object_content [
         a viso-am:sampling_algorithm
      ];
      osmo:has_aspect_text_content
         "Well-tempered metadynamics"
   ].
\end{verbatim}

\subsubsection{Taxonomy of physical equations and relation between OSMO and \newydd{the Review of Materials Modelling (}RoMM\newydd{)}}
\label{subsubsec:physical-equations}

In OSMO, building on the terminology from RoMM,\citep{DeBaas17}
common PEs in materials modelling are classified into 25
types, represented by subclasses of the OSMO class \ttl{physi\-cal\_equa\-tion\_type},
at four granularity levels (instances of the OSMO class \ttl{granu\-larity\_level}),
\cf~Tab.~\ref{tab:pe-types}.
The characterization of model granularity follows \citet{DeBaas17}
where the scope of each of the RoMM vocabulary categories is discussed in great detail.

Accordingly, particle-based methods are defined to
be \textit{atomistic} if the particles represent single atoms
and \textit{mesoscopic} if they represent multiple
atoms; by this categorization,\citep{DeBaas17} \eg, molecular
models following the united-atom approach are regarded as mesoscopic.
This distinction between atomistic and mesoscopic PEs, however,
is only based on the role ascribed to the discrete particles;
therefore, the same equations can be applied at both levels.
To ensure that the expressive capacity of OSMO matches that of RoMM, MODA,
and EMMO, it is necessary to differentiate between these two levels.\citep{DeBaas17, SCLHGKWWGPAFCOGFBCW18, GFSG19}
For most purposes, however, this is not a crucial distinction, and they
can be jointly referred to as molecular models.

\begin{table}
   \caption{OSMO physical equation types at four granularity levels on the basis of RoMM\citep{DeBaas17}}
   \label{tab:pe-types}
   \footnotesize
  \begin{tabular}{c|cc|l}
   \hline
     granularity level ~&~ PE type ID ~&~ RoMM no. ~&~ class name and category description \\
   \hline
      \tttl{ELECTRONIC} ~&~
         EL.1 ~&~ 1.1 ~&~ \tttl{pe\_type\_electro\-nic\_qm\_ab\-initio} \\
            ~&~ ~&~ ~&~ ab-initio quantum mechanical and first-principle models \\
      ~&~
         EL.2 ~&~ 1.2 ~&~ \tttl{pe\_type\_electro\-nic\_many\-body\_effec\-tive} \\
            ~&~ ~&~ ~&~ electronic many-body and effective Hamiltonian models \\
      ~&~
         EL.3 ~&~ 1.3 ~&~ \tttl{pe\_type\_electro\-nic\_time\_de\-pendent} \\
            ~&~ ~&~ ~&~ QM modelling of the response to time-dependent fields \\
      ~&~
         EL.4 ~&~ 1.4 ~&~ \tttl{pe\_type\_electro\-nic\_charge\_trans\-port} \\
            ~&~ ~&~ ~&~ statistical charge transport models \\
      ~&~
         EL.5 ~&~ 1.5 ~&~ \tttl{pe\_type\_electro\-nic\_spin\_trans\-port} \\
            ~&~ ~&~ ~&~ statistical electronic spin transport models \\
   \hline
      \tttl{ATOMISTIC} and ~&~
         A.1 ~&~ 2.1 ~&~ \tttl{pe\_type\_atomis\-tic\_densi\-ty\_functio\-nal} \\
      \tttl{MESOSCOPIC}, \ie, ~&~
         M.1 ~&~ 3.1 ~&~ \tttl{pe\_type\_meso\-scopic\_densi\-ty\_func\-tional} \\
            molecular models ~&~ ~&~ ~&~ classical-mechanical DFT \\
      ~&~
         A.2 ~&~ 2.2 ~&~ \tttl{pe\_type\_atomis\-tic\_mole\-cular\_sta\-tics} \\
      ~&~
         M.2 ~&~ --- ~&~ \tttl{pe\_type\_meso\-scopic\_mole\-cular\_sta\-tics} \\
            ~&~ ~&~ ~&~ energy minimization and molecular statics \\
      ~&~
         A.3 ~&~ 2.3 ~&~ \tttl{pe\_type\_atomis\-tic\_mole\-cular\_dyna\-mics} \\
      ~&~
         M.3 ~&~ 3.2 ~&~ \tttl{pe\_type\_meso\-scopic\_mole\-cular\_dyna\-mics} \\
            ~&~ ~&~ ~&~ MD based on classical equations of motion \\
      ~&~
         A.4 ~&~ 2.4 ~&~ \tttl{pe\_type\_atomis\-tic\_parti\-tion\_func\-tion} \\
      ~&~
         M.4 ~&~ 3.3 ~&~ \tttl{pe\_type\_meso\-scopic\_parti\-tion\_func\-tion} \\
            ~&~ ~&~ ~&~ molecular partition-function equations (\eg, for MC) \\
      ~&~
         A.5 ~&~ 2.5 ~&~ \tttl{pe\_type\_atomis\-tic\_spin\_model} \\
      ~&~
         M.5 ~&~ 3.4 ~&~ \tttl{pe\_type\_meso\-scopic\_micro\-magnetism} \\
            ~&~ ~&~ ~&~ atomistic spin models (A.5), micromagnetism models (M.5)  \\
      ~&~
         A.6 ~&~ 2.6, 2.7 ~&~ \tttl{pe\_type\_atomis\-tic\_statis\-tical\_trans\-port} \\
      ~&~
         M.6 ~&~ 3.5 ~&~ \tttl{pe\_type\_meso\-scopic\_statis\-tical\_trans\-port} \\
            ~&~ ~&~ ~&~ molecular-level statistical transport models \\
   \hline
      \tttl{CONTINUUM} ~&~
         CO.1 ~&~ 4.1 ~&~ \tttl{pe\_type\_con\-tinuum\_so\-lid\_mecha\-nics} \\
            ~&~ ~&~ ~&~ continuum solid mechanics \\
      ~&~
         CO.2 ~&~ 4.2 ~&~ \tttl{pe\_type\_con\-tinuum\_flu\-id\_mecha\-nics} \\
            ~&~ ~&~ ~&~ continuum fluid mechanics \\
      ~&~
         CO.3 ~&~ 4.3 ~&~ \tttl{pe\_type\_con\-tinuum\_heat\_trans\-fer} \\
            ~&~ ~&~ ~&~ thermomechanics and continuum modelling of heat transfer \\
      ~&~
         CO.4 ~&~ 4.4.2 ~&~ \tttl{pe\_type\_con\-tinuum\_phase\_field} \\
            ~&~ ~&~ ~&~ phase field models and density gradient theory \\
      ~&~
         CO.5 ~&~ 4.4.1 ~&~ \tttl{pe\_type\_con\-tinuum\_thermo\-dynamics} \\
            ~&~ ~&~ ~&~ continuum thermodynamics \\
      ~&~
         CO.6 ~&~ 4.5 ~&~ \tttl{pe\_type\_con\-tinuum\_re\-action\_kine\-tics} \\
            ~&~ ~&~ ~&~ continuum modelling of chemical reaction kinetics \\
      ~&~
         CO.7 ~&~ 4.6 ~&~ \tttl{pe\_type\_con\-tinuum\_electro\-magnetism} \\
            ~&~ ~&~ ~&~ continuum electromagnetism models, including optics \\
      ~&~
         CO.8 ~&~ 4.7 ~&~ \tttl{pe\_type\_con\-tinuum\_pro\-cess\_mo\-del} \\
            ~&~ ~&~ ~&~ continuum process models, including flowchart models \\
   \hline
   \end{tabular}
\end{table}

\section{Conclusion}
\label{sec:conclusion}

The ontologies presented in this work, VISO and OSMO, are
intended to play a role as building blocks within a major organized
effort toward full interoperability of methods, tools, and environments
in computational molecular engineering. This is an ongoing development to
which the VIMMP project contributes together with other projects
(\eg, MarketPlace). These efforts are coordinated by discussions
within the EMMC, an organization open to all modellers, end users, and
service providers in the fields of quantum mechanical, molecular,
and continuum simulation. By specifying workflows in terms
of OSMO, workflow \newydd{environments}\marked{~\sout{management systems}} such as the TaLPas WMS become
interoperable with the VIMMP marketplace and environments from other projects
that will be provided at the virtual marketplace frontend.
Substantial future work will be needed to develop solutions for facilitating the data
ingest into OSMO-compliant infra\-structures by providing user-friendly
tools to describe simulation workflows in computational molecular
engineering according to the approach introduced in the present work.


\begin{acknowledgement}
The co-authors M.T.H., G.B., P.C., S.C., M.C., J.E., P.S., M.A.S., I.T.T.,
and W.L.C.\ acknowledge funding from the European Union's Horizon 2020
research and innovation programme under grant agreement no.\ 760907,
\textit{Virtual Materials Marketplace (VIMMP)}, the co-author
V.L.\ from the European Union's Horizon 2020 research and innovation
programme under grant agreements no.\ 686098, \textit{SmartNanoTox:
Smart Tools for Gauging Nano Hazards}, and 731032, \textit{NanoCommons},
and the co-authors C.N., P.N., and J.V.\ from the German Federal Ministry for
Education and Research (BMBF) under grant no.\ 01IH16008B/C/E
(\lq{}B\rq{}: co-author P.N., \lq{}C\rq{}: co-author C.N., \lq{}E\rq{}: co-author J.V.),
\textit{Task-basierte Lastverteilung und Auto-Tuning in der
Partikelsimulation (TaLPas)}. The authors thank the EMMC
for organizing a series of workshops
dedicated to semantic technology, which provided valuable input that
contributed to the developments presented in this work, and they thank
Y.~Bami, A.~Bhave, A.~Duff, \newydd{A.~M.~Elena,} A.~Fiseni, J.~Friis, E.~Ghedini, G.~Goldbeck,
A.~Hashibon, P.~Klein, R.~Kunze, M.~Lisal, \newydd{A.~Lister,} S.~Metz, S.~Pa\v{r}ez, B.~Plankov\'a,
G.~J.~Schmitz, K.~Sen, A.~Simperler, S.~Stephan, G.~Summer, and C.~Yong
for fruitful discussions.
\end{acknowledgement}

\providecommand{\latin}[1]{#1}
\makeatletter
\providecommand{\doi}
  {\begingroup\let\do\@makeother\dospecials
  \catcode`\{=1 \catcode`\}=2 \doi@aux}
\providecommand{\doi@aux}[1]{\endgroup\texttt{#1}}
\makeatother
\providecommand*\mcitethebibliography{\thebibliography}
\csname @ifundefined\endcsname{endmcitethebibliography}
  {\let\endmcitethebibliography\endthebibliography}{}


\begin{mcitethebibliography}{96}
\providecommand*\natexlab[1]{#1}
\providecommand*\mciteSetBstSublistMode[1]{}
\providecommand*\mciteSetBstMaxWidthForm[2]{}
\providecommand*\mciteBstWouldAddEndPuncttrue
  {\def\EndOfBibitem{\unskip.}}
\providecommand*\mciteBstWouldAddEndPunctfalse
  {\let\EndOfBibitem\relax}
\providecommand*\mciteSetBstMidEndSepPunct[3]{}
\providecommand*\mciteSetBstSublistLabelBeginEnd[3]{}
\providecommand*\EndOfBibitem{}
\mciteSetBstSublistMode{f}
\mciteSetBstMaxWidthForm{subitem}{(\alph{mcitesubitemcount})}
\mciteSetBstSublistLabelBeginEnd
  {\mcitemaxwidthsubitemform\space}
  {\relax}
  {\relax}

\bibitem[M\"oller and Fischer(1990)M\"oller, and Fischer]{MF90}
M\"oller,~D.; Fischer,~J. Vapour liquid equilibrium of a pure fluid from test
  particle method in combination with {$NpT$} molecular dynamics simulations.
  \emph{Mol.\ Phys.} \textbf{1990}, \emph{69}, 463--473\relax
\mciteBstWouldAddEndPuncttrue
\mciteSetBstMidEndSepPunct{\mcitedefaultmidpunct}
{\mcitedefaultendpunct}{\mcitedefaultseppunct}\relax
\EndOfBibitem
\bibitem[Lotfi \latin{et~al.}(1992)Lotfi, Vrabec, and Fischer]{LVF92}
Lotfi,~A.; Vrabec,~J.; Fischer,~J. Vapour liquid equilibria of the
  {L}ennard-{J}ones fluid from the {$NpT$} plus test particle method.
  \emph{Mol.\ Phys.} \textbf{1992}, \emph{76}, 1319\relax
\mciteBstWouldAddEndPuncttrue
\mciteSetBstMidEndSepPunct{\mcitedefaultmidpunct}
{\mcitedefaultendpunct}{\mcitedefaultseppunct}\relax
\EndOfBibitem
\bibitem[Vrabec and Hasse(2002)Vrabec, and Hasse]{VH02}
Vrabec,~J.; Hasse,~H. Grand equilibrium: {V}apour-liquid equilibria by a new
  molecular simulation method. \emph{Mol.\ Phys.} \textbf{2002}, \emph{100},
  3375--3383\relax
\mciteBstWouldAddEndPuncttrue
\mciteSetBstMidEndSepPunct{\mcitedefaultmidpunct}
{\mcitedefaultendpunct}{\mcitedefaultseppunct}\relax
\EndOfBibitem
\bibitem[Deublein \latin{et~al.}(2011)Deublein, Eckl, Stoll, Lishchuk,
  Guevara~Carri{\'o}n, Glass, Merker, Bernreuther, Hasse, and
  Vrabec]{DESLGGMBHV11}
Deublein,~S.; Eckl,~B.; Stoll,~J.; Lishchuk,~S.~V.; Guevara~Carri{\'o}n,~G.;
  Glass,~C.~W.; Merker,~T.; Bernreuther,~M.; Hasse,~H.; Vrabec,~J. {ms2: A}
  molecular simulation tool for thermodynamic properties. \emph{Comp.\ Phys.\
  Comm.} \textbf{2011}, \emph{182}, 2350--2367\relax
\mciteBstWouldAddEndPuncttrue
\mciteSetBstMidEndSepPunct{\mcitedefaultmidpunct}
{\mcitedefaultendpunct}{\mcitedefaultseppunct}\relax
\EndOfBibitem
\bibitem[Glass \latin{et~al.}(2014)Glass, Reiser, Rutkai, Deublein, K{\"o}ster,
  Guevara~Carri{\'o}n, Wafai, Horsch, Bernreuther, Windmann, Hasse, and
  Vrabec]{GRRDKGWHBWHV14}
Glass,~C.~W.; Reiser,~S.; Rutkai,~G.; Deublein,~S.; K{\"o}ster,~A.;
  Guevara~Carri{\'o}n,~G.; Wafai,~A.; Horsch,~M.; Bernreuther,~M.;
  Windmann,~T.; Hasse,~H.; Vrabec,~J. {ms2: A} molecular simulation tool for
  thermodynamic properties, new version release. \emph{Comp.\ Phys.\ Comm.}
  \textbf{2014}, \emph{185}, 3302--3306\relax
\mciteBstWouldAddEndPuncttrue
\mciteSetBstMidEndSepPunct{\mcitedefaultmidpunct}
{\mcitedefaultendpunct}{\mcitedefaultseppunct}\relax
\EndOfBibitem
\bibitem[Rutkai \latin{et~al.}(2017)Rutkai, K{\"o}ster, Guevara~Carri{\'o}n,
  Janzen, Schappals, Glass, Bernreuther, Wafai, Stephan, Kohns, Reiser,
  Deublein, Horsch, Hasse, and Vrabec]{RKGJSGBWSKRDHHV17}
Rutkai,~G.; K{\"o}ster,~A.; Guevara~Carri{\'o}n,~G.; Janzen,~T.; Schappals,~M.;
  Glass,~C.~W.; Bernreuther,~M.; Wafai,~A.; Stephan,~S.; Kohns,~M.; Reiser,~S.;
  Deublein,~S.; Horsch,~M.; Hasse,~H.; Vrabec,~J. {ms2}: {A} molecular
  simulation tool for thermodynamic properties, release 3.0. \emph{Comp.\
  Phys.\ Comm.} \textbf{2017}, \emph{221}, 343--351\relax
\mciteBstWouldAddEndPuncttrue
\mciteSetBstMidEndSepPunct{\mcitedefaultmidpunct}
{\mcitedefaultendpunct}{\mcitedefaultseppunct}\relax
\EndOfBibitem
\bibitem[Vrabec \latin{et~al.}(2001)Vrabec, Stoll, and Hasse]{VSH01}
Vrabec,~J.; Stoll,~J.; Hasse,~H. A set of molecular models for symmetric
  quadrupolar fluids. \emph{J.\ Phys.\ Chem.\ B} \textbf{2001}, \emph{105},
  12126--12133\relax
\mciteBstWouldAddEndPuncttrue
\mciteSetBstMidEndSepPunct{\mcitedefaultmidpunct}
{\mcitedefaultendpunct}{\mcitedefaultseppunct}\relax
\EndOfBibitem
\bibitem[Eckl \latin{et~al.}(2008)Eckl, Vrabec, and Hasse]{EVH08a}
Eckl,~B.; Vrabec,~J.; Hasse,~H. On the application of force fields for
  predicting a wide variety of properties: {E}thylene oxide as an example.
  \emph{Fluid Phase Equilib.} \textbf{2008}, \emph{274}, 16--26\relax
\mciteBstWouldAddEndPuncttrue
\mciteSetBstMidEndSepPunct{\mcitedefaultmidpunct}
{\mcitedefaultendpunct}{\mcitedefaultseppunct}\relax
\EndOfBibitem
\bibitem[Schnabel \latin{et~al.}(2008)Schnabel, Vrabec, and Hasse]{SVH08}
Schnabel,~T.; Vrabec,~J.; Hasse,~H. Molecular simulation study of hydrogen
  bonding mixtures and new molecular models for mono- and dimethylamine.
  \emph{Fluid Phase Equilib.} \textbf{2008}, \emph{263}, 144--159\relax
\mciteBstWouldAddEndPuncttrue
\mciteSetBstMidEndSepPunct{\mcitedefaultmidpunct}
{\mcitedefaultendpunct}{\mcitedefaultseppunct}\relax
\EndOfBibitem
\bibitem[Engin \latin{et~al.}(2011)Engin, Merker, Vrabec, and Hasse]{EMVH11}
Engin,~C.; Merker,~T.; Vrabec,~J.; Hasse,~H. Flexible or rigid molecular
  models? A study on vapour-liquid equilibrium properties of ammonia.
  \emph{Mol.\ Phys.} \textbf{2011}, \emph{109}, 619--624\relax
\mciteBstWouldAddEndPuncttrue
\mciteSetBstMidEndSepPunct{\mcitedefaultmidpunct}
{\mcitedefaultendpunct}{\mcitedefaultseppunct}\relax
\EndOfBibitem
\bibitem[Huang \latin{et~al.}(2011)Huang, Heilig, Hasse, and Vrabec]{HHHV11}
Huang,~Y.-L.; Heilig,~M.; Hasse,~H.; Vrabec,~J. Vapor-liquid equilibria of
  hydrogen chloride, phosgene, benzene, chlorobenzene, ortho-dichlorobenzene,
  and toluene by molecular simulation. \emph{AIChE J.} \textbf{2011},
  \emph{57}, 1043--1060\relax
\mciteBstWouldAddEndPuncttrue
\mciteSetBstMidEndSepPunct{\mcitedefaultmidpunct}
{\mcitedefaultendpunct}{\mcitedefaultseppunct}\relax
\EndOfBibitem
\bibitem[Stephan \latin{et~al.}(2019)Stephan, Horsch, Vrabec, and
   Hasse]{SHVH19}
\newydd{Stephan,~S.; Horsch,~M.~T.; Vrabec,~J.; Hasse,~H. {MolMod -- a}n
   open access database of force fields for molecular simulations of fluids.
   \emph{Mol.\ Sim.} \textbf{2019}, \emph{45}, 806--814}\relax
\mciteBstWouldAddEndPuncttrue
\mciteSetBstMidEndSepPunct{\mcitedefaultmidpunct}
{\mcitedefaultendpunct}{\mcitedefaultseppunct}\relax
\EndOfBibitem
\bibitem[Guevara~Carri{\'o}n \latin{et~al.}(2008)Guevara~Carri{\'o}n,
  Nieto~Draghi, Vrabec, and Hasse]{GNVH08}
Guevara~Carri{\'o}n,~G.; Nieto~Draghi,~C.; Vrabec,~J.; Hasse,~H. Prediction of
  transport properties by molecular simulation: Methanol and ethanol and their
  mixture. \emph{J.\ Phys.\ Chem.\ B} \textbf{2008}, \emph{112},
  16664--16674\relax
\mciteBstWouldAddEndPuncttrue
\mciteSetBstMidEndSepPunct{\mcitedefaultmidpunct}
{\mcitedefaultendpunct}{\mcitedefaultseppunct}\relax
\EndOfBibitem
\bibitem[Huang \latin{et~al.}(2009)Huang, Vrabec, and Hasse]{HVH09}
Huang,~Y.-L.; Vrabec,~J.; Hasse,~H. Prediction of ternary vapor-liquid
  equilibria for 33 systems by molecular simulation. \emph{Fluid Phase
  Equilib.} \textbf{2009}, \emph{287}, 62--69\relax
\mciteBstWouldAddEndPuncttrue
\mciteSetBstMidEndSepPunct{\mcitedefaultmidpunct}
{\mcitedefaultendpunct}{\mcitedefaultseppunct}\relax
\EndOfBibitem
\bibitem[Pa{\v{r}}ez \latin{et~al.}(2013)Pa{\v{r}}ez, Guevara~Carri{\'o}n,
  Hasse, and Vrabec]{PGHV13}
Pa{\v{r}}ez,~S.; Guevara~Carri{\'o}n,~G.; Hasse,~H.; Vrabec,~J. Mutual
  diffusion in the ternary mixture of water + methanol + ethanol and its binary
  subsystems. \emph{Phys.\ Chem.\ Chem.\ Phys.} \textbf{2013}, \emph{15},
  3985--4001\relax
\mciteBstWouldAddEndPuncttrue
\mciteSetBstMidEndSepPunct{\mcitedefaultmidpunct}
{\mcitedefaultendpunct}{\mcitedefaultseppunct}\relax
\EndOfBibitem
\bibitem[Werth \latin{et~al.}(2016)Werth, Kohns, Langenbach, Heilig, Horsch,
  and Hasse]{WKLHHH16}
Werth,~S.; Kohns,~M.; Langenbach,~K.; Heilig,~M.; Horsch,~M.; Hasse,~H.
  Interfacial and bulk properties of vapor-liquid equilibria in the system
  toluene + hydrogen chloride + carbon dioxide by molecular simulation and
  density gradient theory + {PC-SAFT}. \emph{Fluid Phase Equilib.}
  \textbf{2016}, \emph{427}, 219--230\relax
\mciteBstWouldAddEndPuncttrue
\mciteSetBstMidEndSepPunct{\mcitedefaultmidpunct}
{\mcitedefaultendpunct}{\mcitedefaultseppunct}\relax
\EndOfBibitem
\bibitem[Niethammer \latin{et~al.}(2014)Niethammer, Becker, Bernreuther,
  Buchholz, Eckhardt, Heinecke, Werth, Bungartz, Glass, Hasse, Vrabec, and
  Horsch]{NBBBEHWBGHVH14}
Niethammer,~C.; Becker,~S.; Bernreuther,~M.; Buchholz,~M.; Eckhardt,~W.;
  Heinecke,~A.; Werth,~S.; Bungartz,~H.-J.; Glass,~C.~W.; Hasse,~H.;
  Vrabec,~J.; Horsch,~M. {ls1 mardyn: T}he massively parallel molecular
  dynamics code for large systems. \emph{J.\ Chem.\ Theory Comput.}
  \textbf{2014}, \emph{10}, 4455--4464\relax
\mciteBstWouldAddEndPuncttrue
\mciteSetBstMidEndSepPunct{\mcitedefaultmidpunct}
{\mcitedefaultendpunct}{\mcitedefaultseppunct}\relax
\EndOfBibitem
\bibitem[Eckhardt \latin{et~al.}(2013)Eckhardt, Heinecke, Bader, Brehm, Hammer,
  Huber, Kleinhenz, Vrabec, Hasse, Horsch, Bernreuther, Glass, Niethammer,
  Bode, and Bungartz]{EHBBHHKVHHBGNBB13}
Eckhardt,~W.; Heinecke,~A.; Bader,~R.; Brehm,~M.; Hammer,~N.; Huber,~H.;
  Kleinhenz,~H.-G.; Vrabec,~J.; Hasse,~H.; Horsch,~M.; Bernreuther,~M.;
  Glass,~C.~W.; Niethammer,~C.; Bode,~A.; Bungartz,~H.-J. 591 {TFLOPS}
  multi-trillion particles simulation on {SuperMUC}. Supercomputing -- 28th
  International Supercomputing Conference (ISC 2013). Springer: Heidelberg,
  \textbf{2013}; pp 1--12\relax
\mciteBstWouldAddEndPuncttrue
\mciteSetBstMidEndSepPunct{\mcitedefaultmidpunct}
{\mcitedefaultendpunct}{\mcitedefaultseppunct}\relax
\EndOfBibitem
\bibitem[Tchipev \latin{et~al.}(2019)Tchipev, Seckler, Heinen, Vrabec, Gratl,
  Horsch, Bernreuther, Glass, Niethammer, Hammer, Krischok, Resch,
  Kranzlm{\"u}ller, Hasse, Bungartz, and Neumann]{TSHVGHBGNHKRKHBN19}
Tchipev,~N.\newydd{; Seckler,~S.; Heinen,~M.; Vrabec,~J.; Gratl,~F.;
  Horsch,~M.; Bernreuther,~M.; Glass,~C.~W.; Niethammer,~C.; Hammer,~N.;
  Krischok,~B.; Resch,~M.; Kranzlm\"uller,~D.; Hasse,~H.; Bungartz,~H.-J.;
  Neumann,~P.} {TweTriS: T}wenty trillion-atom simulation. \emph{Int.\ J.\
  HPC Appl.} \textbf{2019}\newydd{, \emph{33}, 838--854}\relax
\mciteBstWouldAddEndPuncttrue
\mciteSetBstMidEndSepPunct{\mcitedefaultmidpunct}
{\mcitedefaultendpunct}{\mcitedefaultseppunct}\relax
\EndOfBibitem
\bibitem[Asprion \latin{et~al.}(2014)Asprion, Benfer, Blagov, B{\"o}ttcher,
  Bortz, Welke, Burger, {v}on Harbou, K{\"u}fer, and Hasse]{AB4WBHKH14}
Asprion,~N.; Benfer,~R.; Blagov,~S.; B{\"o}ttcher,~R.; Bortz,~M.; Welke,~R.;
  Burger,~J.; {v}on Harbou,~E.; K{\"u}fer,~K.-H.; Hasse,~H. {INES: I}nterface
  between experiments and simulation. \emph{Comp.\ Aided Chem.\ Eng.}
  \textbf{2014}, \emph{33}, 1159--1164\relax
\mciteBstWouldAddEndPuncttrue
\mciteSetBstMidEndSepPunct{\mcitedefaultmidpunct}
{\mcitedefaultendpunct}{\mcitedefaultseppunct}\relax
\EndOfBibitem
\bibitem[Asprion \latin{et~al.}(2015)Asprion, Benfer, Blagov, B{\"o}ttcher,
  Bortz, Berezhnyi, Burger, {v}on Harbou, K{\"u}fer, and Hasse]{AB6HKH15}
Asprion,~N.; Benfer,~R.; Blagov,~S.; B{\"o}ttcher,~R.; Bortz,~M.;
  Berezhnyi,~M.; Burger,~J.; {v}on Harbou,~E.; K{\"u}fer,~K.-H.; Hasse,~H.
  {INES: A}n interface between experiments and simulation to support the
  development of robust process designs. \emph{Chem.\ Ing.\ Techn.}
  \textbf{2015}, \emph{87}, 1810--1825\relax
\mciteBstWouldAddEndPuncttrue
\mciteSetBstMidEndSepPunct{\mcitedefaultmidpunct}
{\mcitedefaultendpunct}{\mcitedefaultseppunct}\relax
\EndOfBibitem
\bibitem[Cavalcanti(2019)]{Cavalcanti19}
Cavalcanti,~W.~L. Virtual Materials Marketplace (VIMMP). \textbf{2019};
  \lnk{http://www.vimmp.eu/}, date of access: 25th July 2019\relax
\mciteBstWouldAddEndPuncttrue
\mciteSetBstMidEndSepPunct{\mcitedefaultmidpunct}
{\mcitedefaultendpunct}{\mcitedefaultseppunct}\relax
\EndOfBibitem
\bibitem[Hashibon(2019)]{Hashibon19}
Hashibon,~A. {MarketPlace}. \textbf{2019}; \lnk{http://www.the-marketplace-project.eu/},
  date of access: 25th July 2019\relax
\mciteBstWouldAddEndPuncttrue
\mciteSetBstMidEndSepPunct{\mcitedefaultmidpunct}
{\mcitedefaultendpunct}{\mcitedefaultseppunct}\relax
\EndOfBibitem
\bibitem[Aurich \latin{et~al.}(2016)Aurich, Schneider, Mayer, Kirsch, and
  Hasse]{ASMKH16}
Aurich,~J.; Schneider,~F.; Mayer,~P.; Kirsch,~B.; Hasse,~H.
  {O}berfl{\"a}chenerzeugungs-{M}orphologie-{E}igenschafts-{B}eziehungen.
  \emph{Zeitschr.\ wirtsch.\ Fabrikbetr.} \textbf{2016}, \emph{111},
  213--216\relax
\mciteBstWouldAddEndPuncttrue
\mciteSetBstMidEndSepPunct{\mcitedefaultmidpunct}
{\mcitedefaultendpunct}{\mcitedefaultseppunct}\relax
\EndOfBibitem
\bibitem[Burger and Hasse(2013)Burger, and Hasse]{BH13}
Burger,~J.; Hasse,~H. Multi-objective optimization using reduced models in
  conceptual design of a fuel additive production process. \emph{Chem.\ Eng.\
  Sci.} \textbf{2013}, \emph{99}, 118--126\relax
\mciteBstWouldAddEndPuncttrue
\mciteSetBstMidEndSepPunct{\mcitedefaultmidpunct}
{\mcitedefaultendpunct}{\mcitedefaultseppunct}\relax
\EndOfBibitem
\bibitem[Bortz \latin{et~al.}(2014)Bortz, Burger, Asprion, Blagov,
  B{\"o}ttcher, Nowak, Scheithauer, Welke, K{\"u}fer, and Hasse]{BBABBNSWKH14}
Bortz,~M.; Burger,~J.; Asprion,~N.; Blagov,~S.; B{\"o}ttcher,~R.; Nowak,~U.;
  Scheithauer,~A.; Welke,~R.; K{\"u}fer,~K.-H.; Hasse,~H. Multi-criteria
  optimization in chemical process design and decision support by navigation on
  {P}areto sets. \emph{Comp.\ Chem.\ Eng.} \textbf{2014}, \emph{60},
  354--363\relax
\mciteBstWouldAddEndPuncttrue
\mciteSetBstMidEndSepPunct{\mcitedefaultmidpunct}
{\mcitedefaultendpunct}{\mcitedefaultseppunct}\relax
\EndOfBibitem
\bibitem[Bortz \latin{et~al.}(2017)Bortz, Burger, {v}on {H}arbou, Klein,
  Schwientek, Asprion, B{\"o}ttcher, K{\"u}fer, and Hasse]{BBHKSABKH17}
Bortz,~M.; Burger,~J.; {v}on {H}arbou,~E.; Klein,~M.; Schwientek,~J.;
  Asprion,~N.; B{\"o}ttcher,~R.; K{\"u}fer,~K.-H.; Hasse,~H. Efficient approach
  for calculating {P}areto boundaries under uncertainties in chemical process
  design. \emph{Ind.\ Eng.\ Chem.\ Res.} \textbf{2017}, \emph{56},
  12672--12681\relax
\mciteBstWouldAddEndPuncttrue
\mciteSetBstMidEndSepPunct{\mcitedefaultmidpunct}
{\mcitedefaultendpunct}{\mcitedefaultseppunct}\relax
\EndOfBibitem
\bibitem[Forte \latin{et~al.}(2018)Forte, Burger, Langenbach, Hasse,
   Bortz]{FBLHB18}
\newydd{Forte Serrano,~E.; Burger,~J.; Langenbach,~K.; Hasse,~H.; Bortz,~M.
   Multi-criteria optimization for parameterization of {SAFT}-type equations of
   state for water. \emph{AIChE J.} \textbf{2018}, \emph{62}, 226--237}\relax
\mciteBstWouldAddEndPuncttrue
\mciteSetBstMidEndSepPunct{\mcitedefaultmidpunct}
{\mcitedefaultendpunct}{\mcitedefaultseppunct}\relax
\EndOfBibitem
\bibitem[{v}on Harbou \latin{et~al.}(2017){v}on Harbou, Ryll, Schrabback,
  Bortz, and Hasse]{HRSBH17}
{v}on Harbou,~E.; Ryll,~O.; Schrabback,~M.; Bortz,~M.; Hasse,~H. Reactive
  distillation in a dividing-wall column: {M}odel development, simulation, and
  error analysis. \emph{Chem.\ Ing.\ Techn.} \textbf{2017}, \emph{89},
  1315--1324\relax
\mciteBstWouldAddEndPuncttrue
\mciteSetBstMidEndSepPunct{\mcitedefaultmidpunct}
{\mcitedefaultendpunct}{\mcitedefaultseppunct}\relax
\EndOfBibitem
\bibitem[Forte \latin{et~al.}(2019)Forte, Jirasek, Bortz, Burger, Vrabec, and
  Hasse]{FJBBVH19}
Forte,~E.; Jirasek,~F.; Bortz,~M.; Burger,~J.; Vrabec,~J.; Hasse,~H.
  Digitalization in thermodynamics. \emph{Chem.\ Ing.\ Techn.} \textbf{2019},
  \emph{91}, 201--214\relax
\mciteBstWouldAddEndPuncttrue
\mciteSetBstMidEndSepPunct{\mcitedefaultmidpunct}
{\mcitedefaultendpunct}{\mcitedefaultseppunct}\relax
\EndOfBibitem
\bibitem[Hasse and Lenhard(2017)Hasse, and Lenhard]{HL17}
Hasse,~H.; Lenhard,~J. In \emph{Mathematics as a Tool: Tracing New Roles of
  Mathematics in the Sciences}; Lenhard,~J., Carrier,~M., Eds.; Springer: Cham,
  \textbf{2017}; pp 93--115\relax
\mciteBstWouldAddEndPuncttrue
\mciteSetBstMidEndSepPunct{\mcitedefaultmidpunct}
{\mcitedefaultendpunct}{\mcitedefaultseppunct}\relax
\EndOfBibitem
\bibitem[Lenhard and Hasse(2017)Lenhard, and Hasse]{LH17}
Lenhard,~J.; Hasse,~H. In \emph{Technisches Nichtwissen}; Friedrich,~A.,
  Gehring,~P., Hubig,~C., Kaminski,~A., Nordmann,~A., Eds.; Nomos: Baden-Baden,
  \textbf{2017}; pp 69--84\relax
\mciteBstWouldAddEndPuncttrue
\mciteSetBstMidEndSepPunct{\mcitedefaultmidpunct}
{\mcitedefaultendpunct}{\mcitedefaultseppunct}\relax
\EndOfBibitem
\bibitem[Schembera and Iglezakis(2019)Schembera, and Iglezakis]{SI19}
Schembera,~B.; Iglezakis,~D. In \emph{Metadata and Semantic Research};
  Garoufallou,~E., Sartori,~F., Siatri,~R., Zervas,~M., Eds.; Springer: Cham,
  \textbf{2019}; pp 127--132\relax
\mciteBstWouldAddEndPuncttrue
\mciteSetBstMidEndSepPunct{\mcitedefaultmidpunct}
{\mcitedefaultendpunct}{\mcitedefaultseppunct}\relax
\EndOfBibitem
\bibitem[M\"uhleisen and Jentzsch(2011)M\"uhleisen, and Jentzsch]{MJ11}
\newydd{M\"uhleisen,~H.; Jentzsch,~A. In \emph{WWW2011 Workshop on Linked Data
  on the Web (LDOW)}; Bizer,~C., Heath,~T., Berners-Lee,~T., Hausenblas,~M.,
  Eds.; CEUR-WS: Aachen, \textbf{2011}; p 3}\relax
\mciteBstWouldAddEndPuncttrue
\mciteSetBstMidEndSepPunct{\mcitedefaultmidpunct}
{\mcitedefaultendpunct}{\mcitedefaultseppunct}\relax
\EndOfBibitem
\bibitem[Bicarregui(2016)Bicarregui]{Bicarregui16}
\newydd{Bicarregui,~J. Building and sustaining data infrastructures: {P}utting
  policy into practice; \textbf{2016},
  doi:10.6084/m9.figshare.4055538.v2}\relax
\mciteBstWouldAddEndPuncttrue
\mciteSetBstMidEndSepPunct{\mcitedefaultmidpunct}
{\mcitedefaultendpunct}{\mcitedefaultseppunct}\relax
\EndOfBibitem
\bibitem[Mons(2018)Mons]{Mons18}
\newydd{Mons,~B. Data Stewardship for Open Science. CRC: Boca Raton, USA,
  \textbf{2018}}\relax
\mciteBstWouldAddEndPuncttrue
\mciteSetBstMidEndSepPunct{\mcitedefaultmidpunct}
{\mcitedefaultendpunct}{\mcitedefaultseppunct}\relax
\EndOfBibitem
\bibitem[Merkys \latin{et~al.}(2017)Merkys, Mounet, Cepellotti, Marzari,
  Gra{\v{z}}ulis, and Pizzi]{MMCMGP17}
Merkys,~A.; Mounet,~N.; Cepellotti,~A.; Marzari,~N.; Gra{\v{z}}ulis,~S.;
  Pizzi,~G. A posteriori metadata from automated provenance tracking:
  {I}ntegration of {AiiDA} and {TCOD}. \emph{J.\ Cheminformat.} \textbf{2017},
  \emph{9}, 56\relax
\mciteBstWouldAddEndPuncttrue
\mciteSetBstMidEndSepPunct{\mcitedefaultmidpunct}
{\mcitedefaultendpunct}{\mcitedefaultseppunct}\relax
\EndOfBibitem
\bibitem[Zelm \latin{et~al.}(2018)Zelm, Jaekel, Doumeingts, and
  Wollschlaeger]{ZJDW18}
\newydd{Zelm,~M.; Jaekel,~F.-W.; Doumeingts,~G.; Wollschlaeger,~M. Enterprise
  Interoperability. Wiley: London, \textbf{2018}}\relax
\mciteBstWouldAddEndPuncttrue
\mciteSetBstMidEndSepPunct{\mcitedefaultmidpunct}
{\mcitedefaultendpunct}{\mcitedefaultseppunct}\relax
\EndOfBibitem
\bibitem[Lehne \latin{et~al.}(2019)Lehne, Sass, Essenwanger, Schepers, and
  Thun]{LSEST19}
\newydd{Lehne,~M.; Sass,~J.; Essenwanger,~A.; Schepers,~J.; Thun,~S. Why
  digital medicine depends on interoperability. \emph{npj Digit.\ Med.}
  \textbf{2019}, \emph{2}, 79}\relax
\mciteBstWouldAddEndPuncttrue
\mciteSetBstMidEndSepPunct{\mcitedefaultmidpunct}
{\mcitedefaultendpunct}{\mcitedefaultseppunct}\relax
\EndOfBibitem
\bibitem[M{\"u}hleisen \latin{et~al.}(2011)M{\"u}hleisen, Walther, and
  Tolksdorf]{MWT11}
M{\"u}hleisen,~H.; Walther,~T.; Tolksdorf,~R. A survey on self-organized
  semantic storage. \emph{Internat.\ J.\ Web Informat.\ Syst.} \textbf{2011},
  \emph{7}, 205--222\relax
\mciteBstWouldAddEndPuncttrue
\mciteSetBstMidEndSepPunct{\mcitedefaultmidpunct}
{\mcitedefaultendpunct}{\mcitedefaultseppunct}\relax
\EndOfBibitem
\bibitem[Heidorn(2008)]{Heidorn08}
Heidorn,~P.~B. Shedding light on the dark data in the long tail of science.
  \emph{Libr.\ Trends} \textbf{2008}, \emph{57}, 280--299\relax
\mciteBstWouldAddEndPuncttrue
\mciteSetBstMidEndSepPunct{\mcitedefaultmidpunct}
{\mcitedefaultendpunct}{\mcitedefaultseppunct}\relax
\EndOfBibitem
\bibitem[Morbach \latin{et~al.}(2008)Morbach, Wiesner, and Marquardt]{MWM08}
Morbach,~J.; Wiesner,~A.; Marquardt,~W. Onto {CAPE} 2.0: {A} (re-)usable
  ontology for computer-aided process engineering. \emph{Comp.\ Aided Chem.\
  Eng.} \textbf{2008}, \emph{25}, 991--996\relax
\mciteBstWouldAddEndPuncttrue
\mciteSetBstMidEndSepPunct{\mcitedefaultmidpunct}
{\mcitedefaultendpunct}{\mcitedefaultseppunct}\relax
\EndOfBibitem
\bibitem[Allemang and Hendler(2011)Allemang, and Hendler]{AH11}
Allemang,~D.; Hendler,~J. \emph{Semantic Web for the Working Ontologist}, 2nd
  ed.; Morgan Kaufmann: Waltham, USA, \textbf{2011}\relax
\mciteBstWouldAddEndPuncttrue
\mciteSetBstMidEndSepPunct{\mcitedefaultmidpunct}
{\mcitedefaultendpunct}{\mcitedefaultseppunct}\relax
\EndOfBibitem
\bibitem[{Mc}~{G}urk \latin{et~al.}(2017){Mc}~{G}urk, Abela, and
  Debattista]{MAD17}
{Mc}~{G}urk,~S.; Abela,~C.; Debattista,~J. Towards ontology quality assessment.
  {MEPDaW-LDQ} 2017 Joint Proceedings. {CEUR-WS}: Aachen, \textbf{2017}; pp
  94--106\relax
\mciteBstWouldAddEndPuncttrue
\mciteSetBstMidEndSepPunct{\mcitedefaultmidpunct}
{\mcitedefaultendpunct}{\mcitedefaultseppunct}\relax
\EndOfBibitem
\bibitem[Li \latin{et~al.}(2019)Li, Armiento, and Lambrix]{LAL19}
\newydd{Li,~H.; Armiento,~R.; Lambrix,~P. A method for extending ontologies
   with application to the materials science domain. \emph{Data Sci.\ J.}
   \textbf{2019}, \emph{18}, 50}\relax
\mciteBstWouldAddEndPuncttrue
\mciteSetBstMidEndSepPunct{\mcitedefaultmidpunct}
{\mcitedefaultendpunct}{\mcitedefaultseppunct}\relax
\EndOfBibitem
\bibitem[Pizzia \latin{et~al.}(2016)Pizzia, Cepellotti, Sabatini, Marzari, and
  Kozinsky]{PCSMK16}
Pizzia,~G.; Cepellotti,~A.; Sabatini,~R.; Marzari,~N.; Kozinsky,~B. {AiiDA:
  A}utomated interactive infrastructure and database for computational science.
  \emph{Comp.\ Mat.\ Sci.} \textbf{2016}, \emph{111}, 218--230\relax
\mciteBstWouldAddEndPuncttrue
\mciteSetBstMidEndSepPunct{\mcitedefaultmidpunct}
{\mcitedefaultendpunct}{\mcitedefaultseppunct}\relax
\EndOfBibitem
\bibitem[Ribes and Caremoli(2007)Ribes, and Caremoli]{RC07}
Ribes,~A.; Caremoli,~C. Salom{\'e} platform component model for numerical
  simulation. 31st Annual International Computer Software and Applications
  Conference, COMPSAC 2007. IEEE Computer Society: Los Alamitos, USA, \textbf{2007}; pp
  553--564\relax
\mciteBstWouldAddEndPuncttrue
\mciteSetBstMidEndSepPunct{\mcitedefaultmidpunct}
{\mcitedefaultendpunct}{\mcitedefaultseppunct}\relax
\EndOfBibitem
\bibitem[Brush \latin{et~al.}(2016)Brush, Shefchek, and Haendel]{BSH16}
Brush,~M.~H.; Shefchek,~K.; Haendel,~M. {SEPIO: A} semantic model for the
  integration and analysis of scientific evidence. Proceedings of the Joint
  International Conference on Biological Ontology and BioCreative. CEUR-WS:
  Aachen, \textbf{2016}\relax
\mciteBstWouldAddEndPuncttrue
\mciteSetBstMidEndSepPunct{\mcitedefaultmidpunct}
{\mcitedefaultendpunct}{\mcitedefaultseppunct}\relax
\EndOfBibitem
\bibitem[Hastings \latin{et~al.}(2015)Hastings, Jeliazkova, Owen, Tsiliki,
  Munteanu, Steinbeck, and Willighagen]{HJOTMSW15}
Hastings,~J.; Jeliazkova,~N.; Owen,~G.; Tsiliki,~G.; Munteanu,~C.~R.;
  Steinbeck,~C.; Willighagen,~E. {eNanoMapper: H}arnessing ontologies to enable
  data integration for nanomaterial risk assessment. \emph{J.\ Biomed.\
  Semantics} \textbf{2015}, \emph{6}, 10\relax
\mciteBstWouldAddEndPuncttrue
\mciteSetBstMidEndSepPunct{\mcitedefaultmidpunct}
{\mcitedefaultendpunct}{\mcitedefaultseppunct}\relax
\EndOfBibitem
\bibitem[Burger \latin{et~al.}(2017)Burger, Asprion, Blagov, and Bortz]{BABB17}
Burger,~J.; Asprion,~N.; Blagov,~S.; Bortz,~M. Simple perturbation scheme to
  consider uncertainty in equations of state for the use in process simulation.
  \emph{J.\ Chem.\ Eng.\ Data} \textbf{2017}, \emph{62}, 268--274\relax
\mciteBstWouldAddEndPuncttrue
\mciteSetBstMidEndSepPunct{\mcitedefaultmidpunct}
{\mcitedefaultendpunct}{\mcitedefaultseppunct}\relax
\EndOfBibitem
\bibitem[Schappals \latin{et~al.}(2017)Schappals, Mecklenfeld, Kr{\"o}ger,
  Botan, K{\"o}ster, Stephan, Garc{\'i}a, Rutkai, Raabe, Klein, Leonhard,
  Glass, Lenhard, Vrabec, and Hasse]{SMKBKSGRRKLGLVH17}
Schappals,~M.; Mecklenfeld,~A.; Kr{\"o}ger,~L.; Botan,~V.; K{\"o}ster,~A.;
  Stephan,~S.; Garc{\'i}a,~E.~J.; Rutkai,~G.; Raabe,~G.; Klein,~P.;
  Leonhard,~K.; Glass,~C.~W.; Lenhard,~J.; Vrabec,~J.; Hasse,~H. Round robin
  study: Molecular simuation of thermodynamic properties from models with
  internal degrees of freedom. \emph{J.\ Chem.\ Theory Comput.} \textbf{2017},
  \emph{13}, 4270--4280\relax
\mciteBstWouldAddEndPuncttrue
\mciteSetBstMidEndSepPunct{\mcitedefaultmidpunct}
{\mcitedefaultendpunct}{\mcitedefaultseppunct}\relax
\EndOfBibitem
\bibitem[Shudler \latin{et~al.}(2019)Shudler, Vrabec, and Wolf]{SVW19}
Shudler,~S.; Vrabec,~J.; Wolf,~F. Understanding the scalability of molecular
  simulation using empirical performance modeling. Programming and Performance
  Visualization Tools. Springer: Heidelberg, \textbf{2019}; pp 125--143\relax
\mciteBstWouldAddEndPuncttrue
\mciteSetBstMidEndSepPunct{\mcitedefaultmidpunct}
{\mcitedefaultendpunct}{\mcitedefaultseppunct}\relax
\EndOfBibitem
\bibitem[Chalk and Elena(2019)Chalk, and Elena]{CE19}
Chalk,~A. B.~G.; Elena,~A.~M. Task-based parallelism with {OpenMP: A} case
  study with {DL\_POLY\_4}. \emph{Mol.\ Sim.} \textbf{2019},
  doi:10.1080/08927022.2019.1606424\relax
\mciteBstWouldAddEndPuncttrue
\mciteSetBstMidEndSepPunct{\mcitedefaultmidpunct}
{\mcitedefaultendpunct}{\mcitedefaultseppunct}\relax
\EndOfBibitem
\bibitem[Belaud and Pons(2002)Belaud, and Pons]{BP02}
Belaud,~J.-P.; Pons,~M. Open software architecture for process simulation:
  {T}he current status of {CAPE-OPEN} standard. \emph{Comp.\ Aided Chem.\ Eng.}
  \textbf{2002}, \emph{10}, 847--852\relax
\mciteBstWouldAddEndPuncttrue
\mciteSetBstMidEndSepPunct{\mcitedefaultmidpunct}
{\mcitedefaultendpunct}{\mcitedefaultseppunct}\relax
\EndOfBibitem
\bibitem[Belaud and Pons(2014)Belaud, and Pons]{BP14}
Belaud,~J.-P.; Pons,~M. {CAPE-OPEN: I}nteroperability in industrial flowsheet
  simulation software. \emph{Chem.\ Ing.\ Techn.} \textbf{2014}, \emph{86},
  1052--1064\relax
\mciteBstWouldAddEndPuncttrue
\mciteSetBstMidEndSepPunct{\mcitedefaultmidpunct}
{\mcitedefaultendpunct}{\mcitedefaultseppunct}\relax
\EndOfBibitem
\bibitem[Lajmi \latin{et~al.}(2009)Lajmi, Cauvin, and Ziane]{LCZ09}
Lajmi,~A.; Cauvin,~S.; Ziane,~M. A software factory for the generation of
  {CAPE-OPEN} compliant process modelling components. \emph{Comp.\ Aided Chem.\
  Eng.} \textbf{2009}, \emph{27}, 207--212\relax
\mciteBstWouldAddEndPuncttrue
\mciteSetBstMidEndSepPunct{\mcitedefaultmidpunct}
{\mcitedefaultendpunct}{\mcitedefaultseppunct}\relax
\EndOfBibitem
\bibitem[Koo \latin{et~al.}(2017)Koo, Trokanas, and Cecelja]{KTC17}
Koo,~L.; Trokanas,~N.; Cecelja,~F. A semantic framework for enabling model
  integration for biorefining. \emph{Comput.\ Chem.\ Eng.} \textbf{2017},
  \emph{100}, 219--231\relax
\mciteBstWouldAddEndPuncttrue
\mciteSetBstMidEndSepPunct{\mcitedefaultmidpunct}
{\mcitedefaultendpunct}{\mcitedefaultseppunct}\relax
\EndOfBibitem
\bibitem[{D}e {B}aas(2017)]{DeBaas17}
{D}e {B}aas,~A.~F., Ed. \emph{What Makes a Material Function? {L}et me Compute
  the Ways}; EU Publications Office: Luxembourg, \textbf{2017}\relax
\mciteBstWouldAddEndPuncttrue
\mciteSetBstMidEndSepPunct{\mcitedefaultmidpunct}
{\mcitedefaultendpunct}{\mcitedefaultseppunct}\relax
\EndOfBibitem
\bibitem[SCL(2018)]{SCLHGKWWGPAFCOGFBCW18}
\emph{Materials modelling: {T}erminology, classification and metadata}; {CEN}
  workshop agreement 17284, \textbf{2018}\relax
\mciteBstWouldAddEndPuncttrue
\mciteSetBstMidEndSepPunct{\mcitedefaultmidpunct}
{\mcitedefaultendpunct}{\mcitedefaultseppunct}\relax
\EndOfBibitem
\bibitem[Gol(2019)]{Goldbeck19}
Taxonda dashboard. \textbf{2019}; \lnk{http://emmc.info/taxonda-dashboard/}, date of
  access: 25th July 2019\relax
\mciteBstWouldAddEndPuncttrue
\mciteSetBstMidEndSepPunct{\mcitedefaultmidpunct}
{\mcitedefaultendpunct}{\mcitedefaultseppunct}\relax
\EndOfBibitem
\bibitem[Ghedini \latin{et~al.}(2019)Ghedini, Friis, Schmitz, and
  Goldbeck]{GFSG19}
Ghedini,~E.; Friis,~J.; Schmitz,~G.~J.; Goldbeck,~G. European Materials {\&}
  Modelling Ontology (EMMO). \textbf{2019}; \lnk{http://github.com/emmo-repo/EMMO/},
  date of access: 25th July 2019\relax
\mciteBstWouldAddEndPuncttrue
\mciteSetBstMidEndSepPunct{\mcitedefaultmidpunct}
{\mcitedefaultendpunct}{\mcitedefaultseppunct}\relax
\EndOfBibitem
\bibitem[Smith \latin{et~al.}(2016)Smith, Katz, and Niemeyer]{SKN16}
Smith,~A.~M.; Katz,~D.~S.; Niemeyer,~K.~E. Software citation principles.
  \emph{PeerJ Comput.\ Sci.} \textbf{2016}, \emph{2}, e86\relax
\mciteBstWouldAddEndPuncttrue
\mciteSetBstMidEndSepPunct{\mcitedefaultmidpunct}
{\mcitedefaultendpunct}{\mcitedefaultseppunct}\relax
\EndOfBibitem
\bibitem[Katz \latin{et~al.}(2019)Katz, Bouquin, Chue~Hong, Hausman, Jones,
  Chivvis, Clark, Crosas, Druskat, Fenner, Gillespie, Gonz{\'a}lez~Beltr{\'a}n,
  Gruenpeter, Habermann, Haines, Harrison, Henneken, Hwang, Jones, Kelly,
  Kennedy, Leinweber, Rios, Robinson, Todorov, Wu, and
  Zhang]{KBCHJC3DFG3H4JKKLRRTWZ19}
Katz,~D.~S.\newydd{; Bouquin,~D.; Chue Hong,~N.~P.; Hausman,~J.; Jones,~C.;
  Chivvis,~D.; Clark,~T.; Crosas,~M.; Druskat,~S.; Fenner,~M.; Gillespie,~T.;
  Gonz{\'a}lez Beltr{\'a}n,~A.; Gruenpeter,~M.; Habermann,~T.; Haines,~R.;
  Harrison,~M.; Henneken,~E.; Hwang,~L.; Jones,~M.~B.; Kelly,~A.~A.;
  Kennedy,~D.~N.; Leinweber,~K.; Rios,~F.; Robinson,~C.~B.; Todorov,~I.~T.;
  Wu,~M.; Zhang,~Q.} \emph{Software Citation Implementation Challenges};
  \textbf{2019}; arXiv:1905.08674 [cs.CY]\relax
\mciteBstWouldAddEndPuncttrue
\mciteSetBstMidEndSepPunct{\mcitedefaultmidpunct}
{\mcitedefaultendpunct}{\mcitedefaultseppunct}\relax
\EndOfBibitem
\bibitem[Cod()]{CodeMeta}
The CodeMeta Project. \textbf{2019}; \lnk{https://codemeta.github.io/}, date of access: 25th
  July 2019\relax
\mciteBstWouldAddEndPuncttrue
\mciteSetBstMidEndSepPunct{\mcitedefaultmidpunct}
{\mcitedefaultendpunct}{\mcitedefaultseppunct}\relax
\EndOfBibitem
\bibitem[CFF()]{CFF}
The Citation File Format (CFF). \textbf{2019}; \lnk{https://citation-file-format.github.io/},
  date of access: 25th July 2019\relax
\mciteBstWouldAddEndPuncttrue
\mciteSetBstMidEndSepPunct{\mcitedefaultmidpunct}
{\mcitedefaultendpunct}{\mcitedefaultseppunct}\relax
\EndOfBibitem
\bibitem[Malone \latin{et~al.}(2014)Malone, Brown, Lister, Ison, Hull,
  Parkinson, and Stevens]{MBLIHPS14}
Malone,~J.; Brown,~A.; Lister,~A.~L.; Ison,~J.; Hull,~D.; Parkinson,~H.;
  Stevens,~R. The software ontology {(SWO): A} resource for reproducibility in
  biomedical data analysis, curation and digital preservation. \emph{J.\
  Biomed.\ Semant.} \textbf{2014}, \emph{5}, 25\relax
\mciteBstWouldAddEndPuncttrue
\mciteSetBstMidEndSepPunct{\mcitedefaultmidpunct}
{\mcitedefaultendpunct}{\mcitedefaultseppunct}\relax
\EndOfBibitem
\bibitem[Gil \latin{et~al.}(2015)Gil, Ratnakar, and Garijo]{GRG15}
Gil,~Y.; Ratnakar,~V.; Garijo,~D. {OntoSoft: C}apturing scientific software
  metadata. Proceedings of the Eighth ACM International Conference on Knowledge
  Capture (K-CAP). ACM: New York, \textbf{2015}; p~32\relax
\mciteBstWouldAddEndPuncttrue
\mciteSetBstMidEndSepPunct{\mcitedefaultmidpunct}
{\mcitedefaultendpunct}{\mcitedefaultseppunct}\relax
\EndOfBibitem
\bibitem[Ervik \latin{et~al.}(2016)Ervik, Mej\'ia, and M\"uller]{EMM16}
\newydd{Ervik,~\AA.; Mej\'ia,~A.; M\"uller,~E.~A. Bottled {SAFT: A} web app
   providing {SAFT-$\gamma$ Mie} force field parameters for thousands of
   molecular fluids. \emph{J.\ Chem.\ Inf.\ Model.} \textbf{2016},
   \emph{56}, 1609--1614}\relax
\mciteBstWouldAddEndPuncttrue
\mciteSetBstMidEndSepPunct{\mcitedefaultmidpunct}
{\mcitedefaultendpunct}{\mcitedefaultseppunct}\relax
\EndOfBibitem
\bibitem[Ervik \latin{et~al.}(2017)Ervik, Jim\'enez, and M\"uller]{EJM17}
\newydd{Ervik,~\AA.; Jim{\'e}nez Serratos,~G.; M\"uller,~E.~A. {raaSAFT: A}
   framework enabling coarse-grained molecular dynamics simulations based on
   the {SAFT-$\gamma$ Mie} force field. \emph{Comp.\ Phys.\ Comm.}
   \textbf{2017}, \emph{212}, 161--179}\relax
\mciteBstWouldAddEndPuncttrue
\mciteSetBstMidEndSepPunct{\mcitedefaultmidpunct}
{\mcitedefaultendpunct}{\mcitedefaultseppunct}\relax
\EndOfBibitem
\bibitem[Bazhirov(2019)]{Bazhirov19}
\newydd{Bazhirov,~T. Data-centric online ecosystem for digital materials
   science. \textbf{2019}, {arXiv}:1902.10838 [cond-mat.mtrl-sci]}\relax
\mciteBstWouldAddEndPuncttrue
\mciteSetBstMidEndSepPunct{\mcitedefaultmidpunct}
{\mcitedefaultendpunct}{\mcitedefaultseppunct}\relax
\EndOfBibitem
\bibitem[Asuncion and {v}an Sunderen(2010)Asuncion, and {v}an Sunderen]{AS10}
Asuncion,~C.~H.; {v}an Sunderen,~M.~J. In \emph{Enterprise Architecture,
  Integration and Interoperability}; Bernus,~P., Doumeingts,~G., Fox,~M., Eds.;
  Springer: Heidelberg, \textbf{2010}; pp 164--175\relax
\mciteBstWouldAddEndPuncttrue
\mciteSetBstMidEndSepPunct{\mcitedefaultmidpunct}
{\mcitedefaultendpunct}{\mcitedefaultseppunct}\relax
\EndOfBibitem
\bibitem[Weidt~Neiva \latin{et~al.}(2016)Weidt~Neiva, David, Braga, and
  Campos]{WDBC16}
Weidt~Neiva,~F.; David,~J. M.~N.; Braga,~R.; Campos,~F. Towards pragmatic
  interoperability to support collaboration: {A} systematic review and mapping
  of the literature. \emph{Informat.\ Software Technol.} \textbf{2016},
  \emph{72}, 137--150\relax
\mciteBstWouldAddEndPuncttrue
\mciteSetBstMidEndSepPunct{\mcitedefaultmidpunct}
{\mcitedefaultendpunct}{\mcitedefaultseppunct}\relax
\EndOfBibitem
\bibitem[Schoop \latin{et~al.}(2006)Schoop, {d}e Moor, and Dietz]{SDD06}
Schoop,~M.; {d}e Moor,~A.; Dietz,~J. The pragmatic web: {A} manifesto.
  \emph{Comm.\ ACM} \textbf{2006}, \emph{49}, 75--76\relax
\mciteBstWouldAddEndPuncttrue
\mciteSetBstMidEndSepPunct{\mcitedefaultmidpunct}
{\mcitedefaultendpunct}{\mcitedefaultseppunct}\relax
\EndOfBibitem
\bibitem[Weidt~Neiva \latin{et~al.}(2016)Weidt~Neiva, David, Braga, Borges, and
  Campos]{WDBBC16}
Weidt~Neiva,~F.; David,~J. M.~N.; Braga,~R.; Borges,~M. R.~S.; Campos,~F.
  {SM2PIA: A} model to support the development of pragmatic interoperability
  requirements. 2016 IEEE 11th International Conference on Global Software
  Engineering (ICGSE). IEEE: New York, \textbf{2016}; pp 119--128\relax
\mciteBstWouldAddEndPuncttrue
\mciteSetBstMidEndSepPunct{\mcitedefaultmidpunct}
{\mcitedefaultendpunct}{\mcitedefaultseppunct}\relax
\EndOfBibitem
\bibitem[Hristova-Bogaerds \latin{et~al.}(2018)Hristova-Bogaerds, Asinari,
  Konchakova, Bergamasco, Marcos~Ramos, Goldbeck, Hoeche, Swang, Schmitz,
  Klein, Kraft, Macio{\l}, Iannone, and {d}e Baas]{HAKBMGHSSKKMID18}
Hristova-Bogaerds,~D.; Asinari,~P.; Konchakova,~N.; Bergamasco,~L.;
  Marcos~Ramos,~A.; Goldbeck,~G.; Hoeche,~D.; Swang,~O.; Schmitz,~G.;
  Klein,~P.; Kraft,~T.; Macio{\l},~P.; Iannone,~M.; {d}e Baas,~A.
  \emph{Translators Guide, version 2}; \textbf{2018};
  \lnk{https://emmc.info/translators-guide-2/}, date of access: 25th July
  2019\relax
\mciteBstWouldAddEndPuncttrue
\mciteSetBstMidEndSepPunct{\mcitedefaultmidpunct}
{\mcitedefaultendpunct}{\mcitedefaultseppunct}\relax
\EndOfBibitem
\bibitem[Schmitz(2016)]{Schmitz16}
Schmitz,~G.~J. Microstructure modeling in integrated computational materials
  engineering ({ICME}) settings: {C}an {HDF5} provide the basis for an emerging
  standard for describing microstructures? \emph{JOM} \textbf{2016}, \emph{68},
  77--83\relax
\mciteBstWouldAddEndPuncttrue
\mciteSetBstMidEndSepPunct{\mcitedefaultmidpunct}
{\mcitedefaultendpunct}{\mcitedefaultseppunct}\relax
\EndOfBibitem
\bibitem[Oberkampf \latin{et~al.}(2018)Oberkampf, Krieg, Senger, Weber, and
  Colsman]{OKSWC18}
Oberkampf,~H.; Krieg,~H.; Senger,~C.; Weber,~T.; Colsman,~W. {A}llotrope {D}ata
  {F}ormat: {S}emantic data management in life sciences. 11th International
  SWAT4HCLS Conference. \textbf{2018}\relax
\mciteBstWouldAddEndPuncttrue
\mciteSetBstMidEndSepPunct{\mcitedefaultmidpunct}
{\mcitedefaultendpunct}{\mcitedefaultseppunct}\relax
\EndOfBibitem
\bibitem[Asprion and Bortz(2018)Asprion, and Bortz]{AB18}
Asprion,~N.; Bortz,~M. Process modeling, simulation and optimization: From
  single solutions to a multitude of solutions to support decision making.
  \emph{Chem.\ Ing.\ Techn.} \textbf{2018}, \emph{90}, 1727--1738\relax
\mciteBstWouldAddEndPuncttrue
\mciteSetBstMidEndSepPunct{\mcitedefaultmidpunct}
{\mcitedefaultendpunct}{\mcitedefaultseppunct}\relax
\EndOfBibitem
\bibitem[De~Leenheer and Christiaens(2007)De~Leenheer, and Christiaens]{DC07}
De~Leenheer,~P.; Christiaens,~S. Mind the gap! {T}ranscending the tunnel view
  on ontology engineering. Proceedings of the 2nd International Conference on
  Pragmatic Web. ACM: New York, \textbf{2007}; pp 75--82\relax
\mciteBstWouldAddEndPuncttrue
\mciteSetBstMidEndSepPunct{\mcitedefaultmidpunct}
{\mcitedefaultendpunct}{\mcitedefaultseppunct}\relax
\EndOfBibitem
\bibitem[Gan(2009)]{Gan09}
Gan,~M. Enterprise isomorphic mapping mechanism: {T}owards ontology
  interoperability in {EIS} development. 2009 IEEE International Conference on
  e-Business Engineering. IEEE Computer Society: Los Alamitos, USA, \textbf{2009}; pp
  340--345\relax
\mciteBstWouldAddEndPuncttrue
\mciteSetBstMidEndSepPunct{\mcitedefaultmidpunct}
{\mcitedefaultendpunct}{\mcitedefaultseppunct}\relax
\EndOfBibitem
\bibitem[Schembera and Dur{\'a}n(2019)Schembera, and Dur{\'a}n]{SD19}
Schembera,~B.; Dur{\'a}n,~J.~M. Dark data as the new challenge for big data
  science and the introduction of the scientific data officer. \emph{Philos.\
  Technol.} \textbf{2019}, 1--13\relax
\mciteBstWouldAddEndPuncttrue
\mciteSetBstMidEndSepPunct{\mcitedefaultmidpunct}
{\mcitedefaultendpunct}{\mcitedefaultseppunct}\relax
\EndOfBibitem
\bibitem[Horridge(2010)Horridge]{Horridge10}
\newydd{Horridge, M. OWLViz. \textbf{2010},
   \lnk{https://protegewiki.stanford.edu/wiki/OWLViz}, date of access: 11th
   November 2019}\relax
\mciteBstWouldAddEndPuncttrue
\mciteSetBstMidEndSepPunct{\mcitedefaultmidpunct}
{\mcitedefaultendpunct}{\mcitedefaultseppunct}\relax
\EndOfBibitem
\bibitem[Ungerer \latin{et~al.}(2000)Ungerer, Beauvais, Delhommelle, Boutin,
  Rousseau, and Fuchs]{UBDBRF00}
Ungerer,~P.; Beauvais,~C.; Delhommelle,~J.; Boutin,~A.; Rousseau,~B.;
  Fuchs,~A.~H. Optimization of the anisotropic united atoms intermolecular
  potential for n-alkanes. \emph{J.\ Chem.\ Phys.} \textbf{2000}, \emph{112},
  5499--5510\relax
\mciteBstWouldAddEndPuncttrue
\mciteSetBstMidEndSepPunct{\mcitedefaultmidpunct}
{\mcitedefaultendpunct}{\mcitedefaultseppunct}\relax
\EndOfBibitem
\bibitem[Jorgensen(1986)]{Jorgensen86}
Jorgensen,~W.~L. Optimized intermolecular potential functions for liquid
  alcohols. \emph{J.\ Phys.\ Chem.} \textbf{1986}, \emph{90}, 1276--1284\relax
\mciteBstWouldAddEndPuncttrue
\mciteSetBstMidEndSepPunct{\mcitedefaultmidpunct}
{\mcitedefaultendpunct}{\mcitedefaultseppunct}\relax
\EndOfBibitem
\bibitem[Jorgensen \latin{et~al.}(1996)Jorgensen, Maxwell, and
  Tirado~Rives]{JMT96}
Jorgensen,~W.~L.; Maxwell,~D.~S.; Tirado~Rives,~J. Development and resting of
  the {OPLS} all-atom force field on conformational energetics and properties
  of organic liquids. \emph{J.\ Am.\ Chem.\ Soc.} \textbf{1996}, \emph{118},
  11225--11236\relax
\mciteBstWouldAddEndPuncttrue
\mciteSetBstMidEndSepPunct{\mcitedefaultmidpunct}
{\mcitedefaultendpunct}{\mcitedefaultseppunct}\relax
\EndOfBibitem
\bibitem[Chen \latin{et~al.}(2001)Chen, Potoff, and Siepmann]{CPS01}
Chen,~B.; Potoff,~J.~J.; Siepmann,~J.~I. {M}onte {C}arlo calculations for
  alcohols and their mixtures with alkanes. {T}ransferable potentials for phase
  equilibria. {5.\ U}nited-atom description of primary, secondary, and tertiary
  alcohols. \emph{J.\ Phys.\ Chem.\ B} \textbf{2001}, \emph{105},
  3093--3104\relax
\mciteBstWouldAddEndPuncttrue
\mciteSetBstMidEndSepPunct{\mcitedefaultmidpunct}
{\mcitedefaultendpunct}{\mcitedefaultseppunct}\relax
\EndOfBibitem
\bibitem[Lampa \latin{et~al.}(2016)Lampa, Alvarsson, and Spjuth]{LAS16}
Lampa,~S.; Alvarsson,~J.; Spjuth,~O. Towards agile large-scale predictive
  modelling in drug discovery with flow-based programming design principles.
  \emph{J.\ Cheminformat.} \textbf{2016}, \emph{8}, 67\relax
\mciteBstWouldAddEndPuncttrue
\mciteSetBstMidEndSepPunct{\mcitedefaultmidpunct}
{\mcitedefaultendpunct}{\mcitedefaultseppunct}\relax
\EndOfBibitem
\bibitem[Beauchemin(2019)]{Beauchemin19}
Beauchemin,~M. Airflow documentation. \textbf{2019}; \lnk{http://airflow.apache.org/},
  date of access: 25th July 2019\relax
\mciteBstWouldAddEndPuncttrue
\mciteSetBstMidEndSepPunct{\mcitedefaultmidpunct}
{\mcitedefaultendpunct}{\mcitedefaultseppunct}\relax
\EndOfBibitem
\bibitem[Jain \latin{et~al.}(2015)Jain, Ong, Chen, Medasani, Qu, Kocher,
  Brafman, Petretto, Rignanese, Hautier, Gunter, and Persson]{JOCMQKBPRHGP15}
Jain,~A.; Ong,~S.~P.; Chen,~W.; Medasani,~B.; Qu,~X.; Kocher,~M.; Brafman,~M.;
  Petretto,~G.; Rignanese,~G.; Hautier,~G.; Gunter,~D.; Persson,~K.~A.
  {FireWorks}: {A} dynamic workflow system designed for high-throughput
  applications. \emph{Concur.\ Computat.\ Pract.\ Exper.} \textbf{2015},
  \emph{27}, 5037--5059\relax
\mciteBstWouldAddEndPuncttrue
\mciteSetBstMidEndSepPunct{\mcitedefaultmidpunct}
{\mcitedefaultendpunct}{\mcitedefaultseppunct}\relax
\EndOfBibitem
\bibitem[Erdmann \latin{et~al.}(2017)Erdmann, Fischer, Fischer, and
  Rieger]{EFFR17}
Erdmann,~M.; Fischer,~B.; Fischer,~R.; Rieger,~M. Design and execution of
  make-like, distributed analyses based on {S}potify's pipelining package
  {L}uigi. \emph{J.\ Phys.: Confer.\ Ser.} \textbf{2017}, \emph{898},
  072047\relax
\mciteBstWouldAddEndPuncttrue
\mciteSetBstMidEndSepPunct{\mcitedefaultmidpunct}
{\mcitedefaultendpunct}{\mcitedefaultseppunct}\relax
\EndOfBibitem
\bibitem[K{\"o}ster and Rahmann(2012)K{\"o}ster, and Rahmann]{KR12}
K{\"o}ster,~J.; Rahmann,~S. Snakemake: {A} scalable bioinformatics workflow
  engine. \emph{Bioinform.} \textbf{2012}, \emph{28}, 2520--2522\relax
\mciteBstWouldAddEndPuncttrue
\mciteSetBstMidEndSepPunct{\mcitedefaultmidpunct}
{\mcitedefaultendpunct}{\mcitedefaultseppunct}\relax
\EndOfBibitem
\bibitem[Robinson \latin{et~al.}(2002)Robinson, Haley, Lermusiaux, and
  Leslie]{RHLL02}
Robinson,~A.~R.; Haley,~P.~J.; Lermusiaux,~P. F.~J.; Leslie,~W.~G. Predictive
  skill, predictive capability and predictability in ocean forecasting. OCEANS
  '02 MTS/IEEE. IEEE: Piscataway, USA, \textbf{2002}; pp 787--794\relax
\mciteBstWouldAddEndPuncttrue
\mciteSetBstMidEndSepPunct{\mcitedefaultmidpunct}
{\mcitedefaultendpunct}{\mcitedefaultseppunct}\relax
\EndOfBibitem
\bibitem[Calotoiu \latin{et~al.}(2016)Calotoiu, Beckingsale, Earl, Hoefler,
  Karlin, Schulz, and Wolf]{CBEHKSW16}
Calotoiu,~A.; Beckingsale,~D.; Earl,~C.~W.; Hoefler,~T.; Karlin,~I.;
  Schulz,~M.; Wolf,~F. Fast multi-parameter performance modeling. 2016 IEEE
  International Conference on Cluster Computing. IEEE Computer Society: Los
  Alamitos, USA, \textbf{2016}; pp 172--181\relax
\mciteBstWouldAddEndPuncttrue
\mciteSetBstMidEndSepPunct{\mcitedefaultmidpunct}
{\mcitedefaultendpunct}{\mcitedefaultseppunct}\relax
\EndOfBibitem
\bibitem[Shudler \latin{et~al.}(2017)Shudler, Calotoiu, Hoefler, and
  Wolf]{SCHW17}
Shudler,~S.; Calotoiu,~A.; Hoefler,~T.; Wolf,~F. Isoefficiency in practice:
  {C}onfiguring and understanding the performance of task-based applications.
  \emph{SIGPLAN Not.} \textbf{2017}, \emph{52}, 131--143\relax
\mciteBstWouldAddEndPuncttrue
\mciteSetBstMidEndSepPunct{\mcitedefaultmidpunct}
{\mcitedefaultendpunct}{\mcitedefaultseppunct}\relax
\EndOfBibitem
\bibitem[Cheong \latin{et~al.}(2018)Cheong, Garijo, Cheung, and Gil]{CGCG18}
\newydd{Cheong,~K.; Garijo,~D.; Cheung,~W.~K.; Gil,~Y. {PSM-Flow:
   P}robabilistic subgraph mining for discovering reusable fragments in
   workflows. 2018 IEEE/WIC/ACM International Conference on Web Intelligence.
   IEEE: Piscataway, USA, \textbf{2018}; pp 166--173}\relax
\mciteBstWouldAddEndPuncttrue
\mciteSetBstMidEndSepPunct{\mcitedefaultmidpunct}
{\mcitedefaultendpunct}{\mcitedefaultseppunct}\relax
\EndOfBibitem
\bibitem[Rutkai and Vrabec(2015)Rutkai, and Vrabec]{RV15}
Rutkai,~G.; Vrabec,~J. Empirical fundamental equation of state for phosgene
  based on molecular simulation data. \emph{J.\ Chem.\ Eng.\ Data}
  \textbf{2015}, \emph{60}, 2895--2905\relax
\mciteBstWouldAddEndPuncttrue
\mciteSetBstMidEndSepPunct{\mcitedefaultmidpunct}
{\mcitedefaultendpunct}{\mcitedefaultseppunct}\relax
\EndOfBibitem
\bibitem[Lustig(2010)]{Lustig11}
Lustig,~R. Direct molecular {NVT} simulation of the isobaric heat capacity,
  speed of sound, and {J}oule-{T}homson coefficient. \emph{Mol.\ Sim.}
  \textbf{2010}, \emph{37}, 457--465\relax
\mciteBstWouldAddEndPuncttrue
\mciteSetBstMidEndSepPunct{\mcitedefaultmidpunct}
{\mcitedefaultendpunct}{\mcitedefaultseppunct}\relax
\EndOfBibitem
\bibitem[Lustig(2012)]{Lustig12}
Lustig,~R. Statistical analogues for fundamental equation of state derivatives.
  \emph{Mol.\ Phys.} \textbf{2012}, \emph{110}, 3041--3052\relax
\mciteBstWouldAddEndPuncttrue
\mciteSetBstMidEndSepPunct{\mcitedefaultmidpunct}
{\mcitedefaultendpunct}{\mcitedefaultseppunct}\relax
\EndOfBibitem
\bibitem[Pathak \latin{et~al.}(2005)Pathak, Caragea, and Honavar]{PCH05}
Pathak,~J.; Caragea,~D.; Honavar,~V.~G. In \emph{Semantic Web and Databases};
  Bussler,~C., Tannen,~V., Fundulaki,~I., Eds.; Springer: Heidelberg, \textbf{2005}; pp
  41--56\relax
\mciteBstWouldAddEndPuncttrue
\mciteSetBstMidEndSepPunct{\mcitedefaultmidpunct}
{\mcitedefaultendpunct}{\mcitedefaultseppunct}\relax
\EndOfBibitem
\bibitem[Rospocher \latin{et~al.}(2014)Rospocher, Ghidini, and Serafini]{RGS14}
Rospocher,~M.; Ghidini,~C.; Serafini,~L. An ontology for the business process
  modelling notation. Formal Ontology in Information Systems: Proceedings of
  the Eighth International Conference. IOS Press: Amsterdam, \textbf{2014}; pp
  133--146\relax
\mciteBstWouldAddEndPuncttrue
\mciteSetBstMidEndSepPunct{\mcitedefaultmidpunct}
{\mcitedefaultendpunct}{\mcitedefaultseppunct}\relax
\EndOfBibitem
\bibitem[Comuzzi(2019)]{Comuzzi19}
Comuzzi,~M. Ant-colony optimisation for path recommendation in business process
  execution. \emph{J.\ Data Semant.} \textbf{2019}, \emph{8}, 113--128\relax
\mciteBstWouldAddEndPuncttrue
\mciteSetBstMidEndSepPunct{\mcitedefaultmidpunct}
{\mcitedefaultendpunct}{\mcitedefaultseppunct}\relax
\EndOfBibitem
\bibitem[Wi{\'s}niewski \latin{et~al.}(2019)Wi{\'s}niewski, Wi{\'s}niewska, and
  Jarnut]{WWJ19}
Wi{\'s}niewski,~R.; Wi{\'s}niewska,~M.; Jarnut,~M. {C}-exact hypergraphs in
  concurrency and sequentiality analyses of cyber-physical systems specified by
  safe {P}etri nets. \emph{IEEE Access} \textbf{2019}, \emph{7}, 13510 --
  13522\relax
\mciteBstWouldAddEndPuncttrue
\mciteSetBstMidEndSepPunct{\mcitedefaultmidpunct}
{\mcitedefaultendpunct}{\mcitedefaultseppunct}\relax
\EndOfBibitem
\bibitem[Ehrig \latin{et~al.}(1999)Ehrig, Engels, Kreowski, and
  Rozenberg]{EEKR99}
Ehrig,~H.; Engels,~G.; Kreowski,~H.; Rozenberg,~G. \emph{Handbook of Graph
  Grammars and Computing by Graph Transformation}; World Scientific: River
  Edge, USA, \textbf{1999}; Vol.~2\relax
\mciteBstWouldAddEndPuncttrue
\mciteSetBstMidEndSepPunct{\mcitedefaultmidpunct}
{\mcitedefaultendpunct}{\mcitedefaultseppunct}\relax
\EndOfBibitem
\bibitem[Corradini and K{\"o}nig(2019)Corradini, and K{\"o}nig]{CKN19}
Corradini,~A.; K{\"o}nig,~B. Specifying graph languages with type graphs.
  \emph{J.\ Logical Algebr.\ Meth.\ Program.} \textbf{2019}, \emph{104},
  176--200\relax
\mciteBstWouldAddEndPuncttrue
\mciteSetBstMidEndSepPunct{\mcitedefaultmidpunct}
{\mcitedefaultendpunct}{\mcitedefaultseppunct}\relax
\EndOfBibitem
\bibitem[Bauderon \latin{et~al.}(2001)Bauderon, M{\'e}tivier, Mosbah, and
  Sellami]{BMMS01}
Bauderon,~M.; M{\'e}tivier,~Y.; Mosbah,~M.; Sellami,~A. Graph relabelling
  systems: {A} tool for encoding, proving, studying and visualizing distributed
  algorithms. \emph{Electr.\ Notes Theor.\ Comp.\ Sci.} \textbf{2001},
  \emph{51}, 93--107\relax
\mciteBstWouldAddEndPuncttrue
\mciteSetBstMidEndSepPunct{\mcitedefaultmidpunct}
{\mcitedefaultendpunct}{\mcitedefaultseppunct}\relax
\EndOfBibitem
\bibitem[Ceusters and Smith(2015)Ceusters, and Smith]{CS15}
Ceusters,~W.; Smith,~B. Aboutness: {T}owards foundations for the information
  artifact ontology. Proceedings of the International Conference on Biomedical
  Ontology. CEUR-WS: Aachen, \textbf{2015}\relax
\mciteBstWouldAddEndPuncttrue
\mciteSetBstMidEndSepPunct{\mcitedefaultmidpunct}
{\mcitedefaultendpunct}{\mcitedefaultseppunct}\relax
\EndOfBibitem
\bibitem[Hodgson \latin{et~al.}(2019)Hodgson, Keller, Hodges, and
  Spivak]{HKHS19}
Hodgson,~R.; Keller,~P.~J.; Hodges,~J.; Spivak,~J. QUDT ontologies. \textbf{2019};
  \lnk{http://www.qudt.org/}, date of access: 25th July 2019\relax
\mciteBstWouldAddEndPuncttrue
\mciteSetBstMidEndSepPunct{\mcitedefaultmidpunct}
{\mcitedefaultendpunct}{\mcitedefaultseppunct}\relax
\EndOfBibitem
\end{mcitethebibliography}
\end{document}